# Quantum image processing?

**Mario Mastriani**

DLQS LLC, 4431 NW 63RD Drive, Coconut Creek, FL 33073, USA
mmastri@gmail.com

*Abstract*—This paper presents a number of problems concerning the practical (real) implementation of the techniques known as Quantum Image Processing. The most serious problem is the recovery of the outcomes after the quantum measurement, which will be demonstrated in this work that is equivalent to a noise measurement, and it is not considered in the literature on the subject. It is noteworthy that this is due to several factors: 1) a classical algorithm that uses Dirac's notation and then it is coded in MATLAB does not constitute a quantum algorithm, 2) the literature emphasizes the internal representation of the image but says nothing about the classical-to-quantum and quantum-to-classical interfaces and how these are affected by decoherence, 3) the literature does not mention how to implement in a practical way (at the laboratory) these proposals internal representations, 4) given that Quantum Image Processing works with generic qubits this requires measurements in all axes of the Bloch sphere, logically, and 5) among others. In return, the technique known as Quantum Boolean Image Processing is mentioned, which works with computational basis states (CBS), exclusively. This methodology allows us to avoid the problem of quantum measurement, which alters the results of the measured except in the case of CBS. Said so far is extended to quantum algorithms outside image processing too.

# 1 Introduction

Quantum computation and quantum information is the study of the information processing tasks that can be accomplished using quantum mechanical systems. Like many simple but profound ideas it was a long time before anybody thought of doing information processing using quantum mechanical systems [1].

Quantum computation is the field that investigates the computational power and other properties of computers based on quantum-mechanical principles. An important objective is to find quantum algorithms that are significantly faster than any classical algorithm solving the same problem. The field started in the early 1980s with suggestions for analog quantum computers by Paul Benioff [2] and Richard Feynman [3, 4], and reached more digital ground when in 1985 David Deutsch defined the universal quantum Turing machine [5]. The following years saw only sparse activity, notably the development of the first algorithms by Deutsch and Jozsa [6] and by Simon [7], and the development of quantum complexity theory by Bernstein and Vazirani [8]. However, interest in the field increased tremendously after Peter Shor's very surprising discovery of efficient quantum algorithms (or simulations on a quantum computer) for the problems of integer factorization and discrete logarithms in 1994 [9].

*Quantum Information Processing (QuIn)* - The main concepts related to Quantum Information Processing may be grouped in the next topics: quantum bit (qubit, which is the elemental quantum information unity), Bloch's Sphere (geometric environment for qubit representation), Hilbert's Space (which generalizes the notion of Euclidean space), Schrödinger's Equation (which is a partial differential equation that describes how the quantum state of a physical system changes with time.), Unitary Operators, Quantum Circuits (in quantum information theory, a quantum circuit is a model for quantum computation in which a computation is a sequence of quantum gates, which are reversible transformations on a quantum mechanical analog of an *n*-bit register. This analogous structure is referred to as an *n*-qubit register.), Quantum Gates (in quantum computing and specifically the quantum circuit model of computation, a quantum gate or quantum logic gate is a basic quantum circuit operating on a small number of qubits), and Quantum Algorithms (in quantum computing, a quantum algorithm is an algorithm which runs on a realistic model of quantum computation, the most commonly used model being the quantum circuit model of computation) [1, 10-12].

Nowadays, other concepts complement our knowledge about QuIn, they are:

*Quantum Signal Processing (QuSP)* - The main idea is to take a classical signal, sample it, quantify it (for example, between 0 and 255), use a classical-to-quantum interface, give an internal representation to that signal, make a processing to that quantum signal (denoising, compression, among others), measure the result, use a quantum-to-classical interface and subsequently detect the classical outcome signal. Interestingly, and as will be seen later, the quantum image processing has aroused more interest than QuSP. In the words of its creator: "many new classes of signal processing algorithms have been developed by emulating the behavior of physical systems. There are also many examples in the signal processing literature in which new classes of algorithms have been developed by artificially imposing physical constraints on implementations that are not inherently subject to these constraints". Therefore, Quantum Signal Processing (QuSP) is a signal processing framework [13, 14] that is aimed at developing new or modifying existing signal processing algorithms by borrowing from the principles of quantum mechanics and some of its interesting axioms and constraints. However, in contrast to such fields as quantum computing and quantum information theory, it does not inherently depend on the physics associated with quantum mechanics. Consequently, in developing the QuSP framework we are free to impose quantum mechanical constraints that we find useful and to avoid those that are not. This framework provides a unifying conceptual structure for a variety of traditional processing techniques and a precise mathematical setting for developing generalizations and extensions of algorithms, leading to a potentially useful paradigm for signal processing with applications in areas including frame theory, quantization and sampling methods, detection, parameter estimation, covariance shaping, and multiuser wireless communication systems." The truth is that to date, papers on this discipline are less than half a dozen, and its practical use is practically nil. Moreover, although it is an interesting idea, developed so far, does not withstand further comment.

***Quantum Image Processing (QuIP)*** - it is a young discipline and it is in training now, however, it's much more developed than QuSP. QuIP starts in 1997. That year, Vlasov proposed a method of using quantum computation to recognize so-called *orthogonal images* [15]. Five years later, in 2002, Schutzhold described a quantum algorithm that searches specific patterns in binary images [16]. A year later, in October 2003, Beach, Lomont, and Cohen from Cybernet Systems Corporation, (an organization with a close cooperative relationship with the US Defense Department) demonstrated the possibility that quantum algorithms (such as Grover's algorithm) can be used in image processing. In that paper, they describe a method which uses a quantum algorithm to detect the posture of certain targets. Their study implies that quantum image processing may, in future, play a valuable role during wartime [17].

Later, we can found the works of Venegas-Andraca [18], where he proposes quantum image representations such as Qubit Lattice [19, 20]; in fact, this is the first doctoral thesis in the specialty, The history continues with the quantum image representation via the Real Ket [21] of Latorre Sentís, with a special interest in image compression in a quantum context. A new stage begins with the proposal of Le et al. [22], for a flexible representation of quantum images to provide a representation for images on quantum computers in the form of a normalized state which captures information about colors and their corresponding positions in the images. History continues up to date by different authors and their innovative internal representation techniques of the image [23-44].

Very similar to the case of QuSP, the idea in back of QuIP is to take a classic image (captured by a digital camera or photon counter) and place it in a quantum machine through a classical-to-quantum interface, give some internal representation to the image using the procedures mentioned above, perform processing on it (denoising, compression, among others), measure the results, restore the image through another interface but this time quantum-classical, y ready. The contribution of a quantum machine over a classic machine when it comes to process images it is that the former has much more power of processing. This last advantage can handle images and algorithms of a high computational cost, which would be unmanageable in a classic machine in a practical sense.

The problem of this discipline lies in its genetic, given that QuIP is the daughter of Quantum Information Processing and Digital Image Processing, thus, we fall into the old dilemma of teaching, i.e.: to teach Latin to Peter, we should know more about Latin and more about Peter? The answer is simple: we should know very well of both, but the mission becomes impossible. In other words, what is acceptable in Quantum Information Processing, and (at the same time) inadmissible in Digital Image Processing?

The mentioned problem begins with the quantum measurement, then, if after processing the image within the quantum computer, we want to retrieve the result by tomography of quantum states, we will encounter a serious obstacle, this is:

*if we make a tomography of quantum states in Quantum Information Processing (even, this can be extended to any method of quantum measurement after the tomography) with an error of 6% in our knowledge of the state, this constitutes an excellent measure of such state, but on the other hand, and this time from the standpoint of Digital Image Processing [45-48], an error of 6 % in each pixel of the outcome image constitutes a disaster, since this error becomes unmanageable and exaggerated noise. So overwhelming is the aforementioned disaster that the recovered image loses its visual intelligibility, i.e., its look and feel, and morphology, due to the destruction of edges and textures.*

This speaks clearly (and for this purpose, one need only read the papers of QuIP cited above) that these works are based on computer simulations in classical machines, exclusively (in most cases in MATLAB® [49]), and do not represent test in a laboratory of Quantum Physics. In fact, if these field trials were held, the result would be the aforementioned. We just have to go to the lab and try with a single pixel of an image, then extrapolate the results to the entire image and therefore the inconvenience will be explicit. On the other hand, today there are obvious difficulties to treat a full image inside a quantum machine, however, there is no difficulty for a single pixel, since that pixel represents a single qubit, and this can be tested in any laboratory in the world, without problems. Therefore, there are no excuses.

Definitely, the problem lies in the hostile relationship between the internal representation of the image (inside quantum machine), the outcome measurement, and the recovery of the image outside of quantum machine. Therefore, the only technique of QuIP that survives is QuBoIP [50]. This is because it works with CBS, exclusively, and the quantum measurement does not affect the value of states. However, it is important to clarify that both, i.e., traditional techniques QuIP and QuBoIP share a common enemy, and this is the decoherence [1, 50].

## 2 Prolegomenous to the problems of Quantum Image Processing

### 2.1 Qubits and Bloch's sphere

The bit is the fundamental concept of classical computation and classical information. Quantum computation and quantum information are built upon an analogous concept, the quantum bit, or qubit for short. In this section we introduce the properties of single and multiple qubits, comparing and contrasting their properties to those of classical bits [1]. The difference between bits and qubits is that a qubit can be in a state other than $|0\rangle$ or $|1\rangle$ [1]. It is also possible to form linear combinations of states, often called superpositions:

$$|\psi\rangle = \alpha|0\rangle + \beta|1\rangle, \qquad (1)$$

where $|\psi\rangle$ is called *wave function*, $|\alpha|^2 + |\beta|^2 = 1$, with the states $|0\rangle$ and $|1\rangle$ are understood as different polarization states of light. Besides, a column vector $|\psi\rangle$ is called a *ket* vector $[\alpha \ \beta]^T$, where, (•)$^T$ means transpose of (•), while a row vector $\langle\psi|$ is called a *bra* vector $[\alpha^* \ \beta^*]$, where, (•)$^*$ means complex conjugate of (•). The numbers $\alpha$ and $\beta$ are complex numbers, although for many purposes not much is lost by thinking of them as real numbers. Put another way, the state of a qubit is a vector in a two-dimensional complex vector space. The special states $|0\rangle$ and $|1\rangle$ are known as Computational Basis States (CBS), and form an orthonormal basis for this vector space, being

$$|0\rangle = \begin{bmatrix} 1 \\ 0 \end{bmatrix} \text{ and } |1\rangle = \begin{bmatrix} 0 \\ 1 \end{bmatrix}$$

One picture useful in thinking about qubits is the following geometric representation. Because $|\alpha|^2 + |\beta|^2 = 1$, we may rewrite Eq.(1) as

$$|\psi\rangle = e^{i\gamma}\left(cos\frac{\theta}{2}|0\rangle + e^{i\phi} sin\frac{\theta}{2}|1\rangle\right) = e^{i\gamma}\left(cos\frac{\theta}{2}|0\rangle + (cos\phi + i\,sin\phi) sin\frac{\theta}{2}|1\rangle\right) \qquad (2)$$

where $0 \leq \theta \leq \pi$, $0 \leq \phi < 2\pi$. We can ignore the factor of $e^{i\gamma}$ out the front, because it has no observable effects [1], and for that reason we can effectively write

$$|\psi\rangle = cos\frac{\theta}{2}|0\rangle + e^{i\phi} sin\frac{\theta}{2}|1\rangle \qquad (3)$$

The numbers $\theta$ and $\phi$ define a point on the unit three-dimensional sphere, as shown in Fig.1.

Quantum mechanics is mathematically formulated in Hilbert space or projective Hilbert space. The space of pure states of a quantum system is given by the one-dimensional subspaces of the corresponding Hilbert space (or the "points" of the projective Hilbert space). In a two-dimensional Hilbert space this is simply the complex projective line, which is a geometrical sphere. This sphere is often called the Bloch's sphere; it provides a useful means of visualizing the state of a single qubit, and often serves as an excellent testbed for ideas about quantum computation and quantum information.

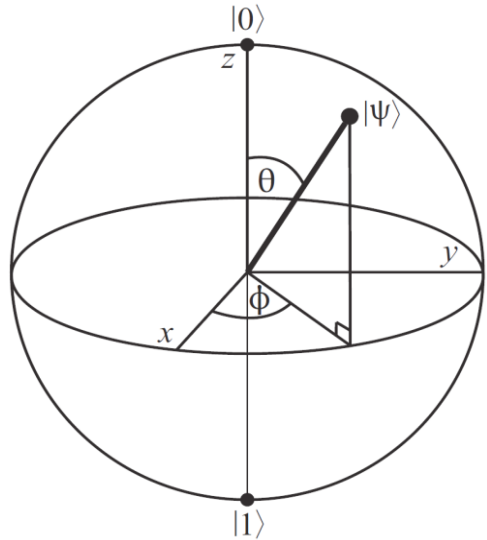

**Fig. 1** Bloch's Sphere.

Many of the operations on single qubits which can be seen in [1] are neatly described within the Bloch's sphere picture. However, it must be kept in mind that this intuition is limited because there is no simple generalization of the Bloch's sphere known for multiple qubits [1]. In the general case, with $n$ qubits there would be $2^n$ possible states [1].

Except in the case where $|\psi\rangle$ is one of the ket vectors $|0\rangle$ or $|1\rangle$ the representation is unique. The parameters $\theta$ and $\phi$, re-interpreted as spherical coordinates, specify a point $\vec{a}=(\sin\theta\cos\phi+\sin\theta\sin\phi+\cos\theta)$ on the unit sphere in $\Re^3$ (according to Eq.2).

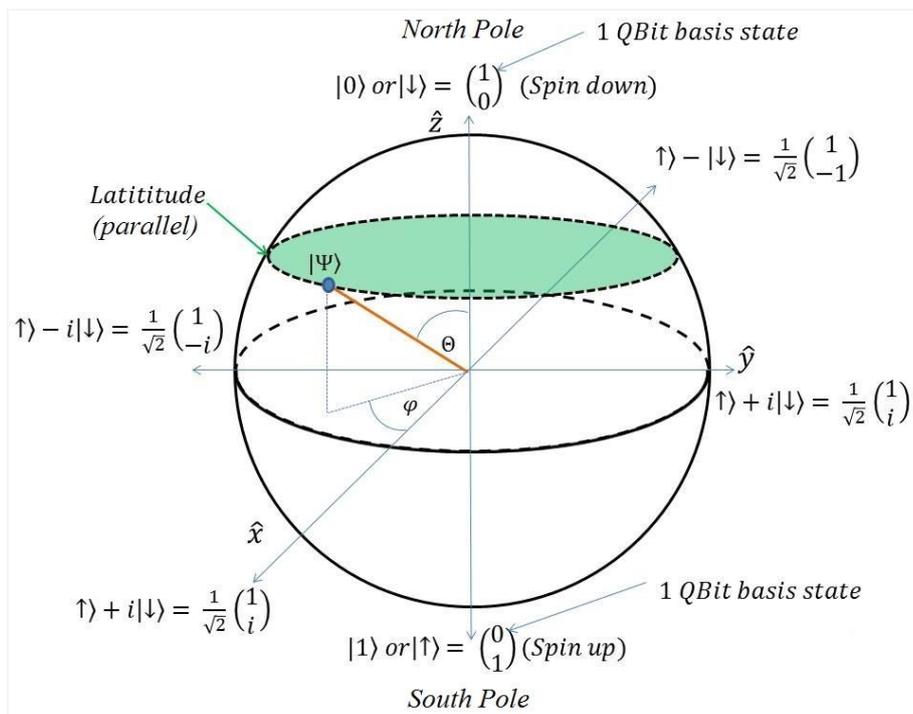

**Fig. 2** Details of the poles, as well as an example of parallel and several qubit states on the sphere.

Figure 2 highlights all components (details) concerning the Bloch's sphere, namely

*Spin down* $= |\downarrow\rangle = |0\rangle = \begin{bmatrix} 1 \\ 0 \end{bmatrix} =$ *qubit basis state = North Pole*

and

*Spin up* $= |\uparrow\rangle = |1\rangle = \begin{bmatrix} 0 \\ 1 \end{bmatrix} =$ *qubit basis state = South Pole*

Both poles play a fundamental role in the development of the quantum computing [1]. Besides, a very important concept to the affections of the development quantum information processing, in general, i.e., the notion of latitude (parallel) on the Bloch's sphere is hinted. Such parallel as shown in green in Fig.2, where we can see the complete coexistence of poles, parallels and meridians on the sphere, including computational basis states $\{|0\rangle, |1\rangle\}$. The poles and the parallels form the geometric bases of criteria and logic needed to implement any quantum gate or circuit.

## 2.2 Schrödinger's equation and unitary operators

A quantum state can be transformed into another state by a unitary operator, symbolized as $U$ ($U: H \rightarrow H$ on a Hilbert space $H$, being called an unitary operator if it satisfies $U^\dagger U = UU^\dagger = I$, where $(\bullet)^\dagger$ is the adjoint of $(\bullet)$, and $I$ is the identity matrix), which is required to preserve inner products: If we transform $|\chi\rangle$ and $|\psi\rangle$ to $U|\chi\rangle$ and $U|\psi\rangle$, then $\langle \chi | UU | \psi \rangle = \langle \chi | \psi \rangle$. In particular, unitary operators preserve lengths:

$$\langle \psi | UU | \psi \rangle = \langle \psi | \psi \rangle = 1, \text{ if } |\psi\rangle \text{ is on the Bloch's sphere (i.e., it is a pure state).} \tag{4}$$

On the other hand, the unitary operator satisfies the following differential equation known as the Schrödinger equation [1, 10-12]:

$$\frac{d}{dt}U(t) = \frac{-i\hat{H}}{\hbar}U(t) \tag{5}$$

where $\hat{H}$ represents the Hamiltonian matrix of the Schrödinger equation, $i = \sqrt[2]{-1}$, and $\hbar$ is the reduced Planck constant, i.e., $\hbar = h/2\pi$. Multiplying both sides of Eq.(5) by $|\psi(0)\rangle$ and setting

$$|\psi(t)\rangle = U(t)|\psi(0)\rangle \tag{6}$$

yields

$$\frac{d}{dt}|\psi(t)\rangle = \frac{-i\hat{H}}{\hbar}|\psi(t)\rangle \tag{7}$$

Equation (6) represents the quantum algorithm employed in Quantum Information Processing. Besides, the solution to the Schrödinger equation is given by the matrix exponential of the Hamiltonian matrix:

$$U(t) = e^{\frac{-i\hat{H}t}{\hbar}} \qquad \text{(if Hamiltonian is not time dependent)} \tag{8}$$

and

$$U(t) = e^{\frac{-i}{\hbar}\int_0^t \hat{H} dt} \qquad \text{(if Hamiltonian is time dependent)} \tag{9}$$

Thus the probability amplitudes evolve across time according to the following equation:

$$|\psi(t)\rangle = e^{\frac{-i\hat{H}t}{\hbar}}|\psi(0)\rangle \quad \text{(if Hamiltonian is not time dependent)} \tag{10}$$

or

$$|\psi(t)\rangle = e^{\frac{-i}{\hbar}\int_0^t \hat{H} dt}|\psi(0)\rangle \quad \text{(if Hamiltonian is time dependent)} \tag{11}$$

The Eq.(10) is the main piece in building circuits, gates and quantum algorithms, being $U$ who represents such elements [1].

Finally, the discrete version of Eq.(7) is

$$|\psi_{n+1}\rangle = \frac{-i\hat{H}}{\hbar}|\psi_n\rangle, \tag{12}$$

for a time dependent (or not) Hamiltonian, being $n$ the discrete time.

### 2.3 Quantum Circuits, Gates, and Algorithms; Reversibility and Quantum Measurement

As we can see in Fig.3, and remember Eq.(6), the quantum algorithm (identical case to circuits and gates) viewed as a transfer (or mapping input-to-output) has two types on output:

a) the result of algorithm (circuit of gate), i.e., $|\psi_n\rangle$
b) part of the input $|\psi_0\rangle$, i.e., $|\underline{\psi_0}\rangle$ (underlined $|\psi_0\rangle$), in order to impart reversibility to the circuit, which is a critical need in quantum computing [1].

Besides, we can see clearly a module for measuring $|\psi_n\rangle$ with their respective output, i.e., $|\psi_n\rangle_{pm}$ (or, $|\psi_n\rangle$ post-measurement), and a number of elements needed for the physical implementation of the quantum algorithm (circuit or gate), namely: control, ancilla and trash [1].

In Fig.3 as well as in the rest of them (unlike [1]) a single fine line represents a wire carrying *1* qubit or *N* qubits (qudit), interchangeably, while a single thick line represents a wire carrying *1* or *N* classical bits, interchangeably too. However, the mentioned concept of reversibility is closely related to energy consumption, and hence to the Landauer's Principle [1].

On the other hand, computational complexity studies the amount of time and space required to solve a computational problem. Another important computational resource is energy. In [1], the authors show the energy requirements for computation. Surprisingly, it turns out that computation, both classical and quantum, can in principle be done without expending any energy! Energy consumption in computation turns out to be deeply linked to the reversibility of the computation. In other words, it is inexcusable the need of the $|\underline{\psi_0}\rangle$ presence to the output of quantum gate [1].

On the other hand, in quantum mechanics, measurement is a non-trivial and highly counter-intuitive process. Firstly, because measurement outcomes are inherently probabilistic, i.e. regardless of the carefulness in the preparation of a measurement procedure, the possible outcomes of such measurement will be distributed according to a certain probability distribution [1]. Secondly, once the measurement has been performed [1], we can see that the very act of measuring alters the measurement, since there is a disturbance, and this is due to the fact that a quantum system is unavoidably altered due to the interaction with the measurement apparatus [50]. Consequently, for an arbitrary quantum system, pre-measurement and post-measurement quantum states are different in general [1].

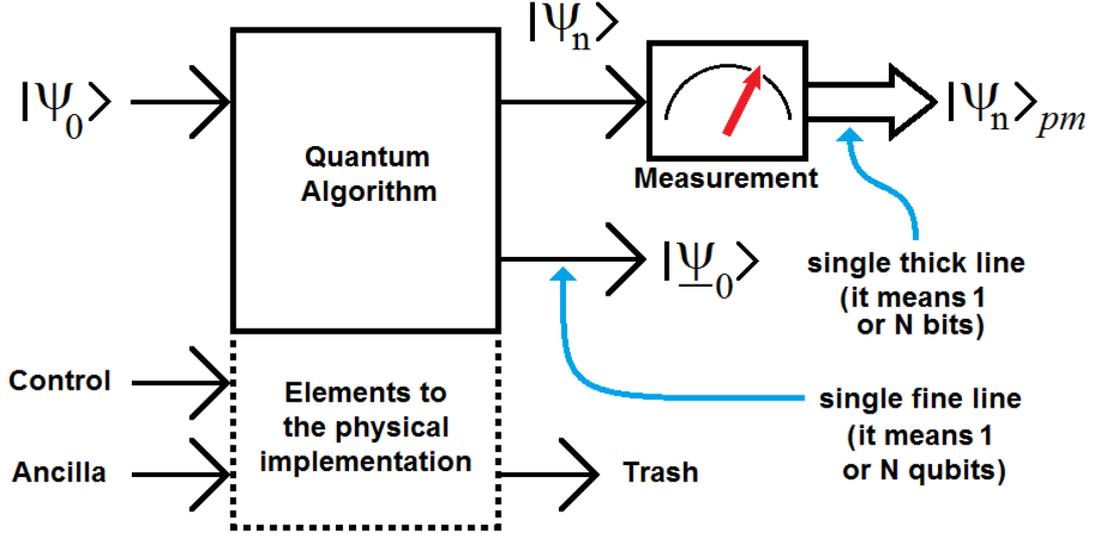

**Fig. 3** Module to measuring, quantum algorithm and the elements needs to its physical implementation.

***Postulate.*** Quantum measurements are described by a set of measurement operators $\{\hat{M}_m\}$, index $m$ labels the different measurement outcomes, which act on the state space of the system being measured. Measurement outcomes correspond to values of *observables*, such as position, energy and momentum, which are Hermitian operators [1] corresponding to physically measurable quantities.

Let $|\psi\rangle$ be the state of the quantum system immediately before the measurement. Then, the probability that result $m$ occurs is given by

$$p(m) = \langle\psi|\hat{M}_m^\dagger \hat{M}_m|\psi\rangle \tag{13}$$

where

$$\hat{M}_m = |\psi_m\rangle\langle\psi_m| \tag{14}$$

and the post-measurement quantum state is

$$|\psi_m\rangle_{pm} = \frac{\hat{M}_m|\psi\rangle}{\sqrt{\langle\psi|\hat{M}_m^\dagger \hat{M}_m|\psi\rangle}} \tag{15}$$

Operators $\hat{M}_m$ must satisfy the completeness relation of Eq.(16), because that guarantees that probabilities will sum to one; see Eq.(17) [1]:

$$\sum_m \hat{M}_m^\dagger \hat{M}_m = I \tag{16}$$

$$\sum_m \langle\psi|\hat{M}_m^\dagger \hat{M}_m|\psi\rangle = \sum_m p(m) = 1 \tag{17}$$

as well as, orthogonality and positive-definite condition, respectively,

$$\hat{M}_i^\dagger \hat{M}_j = 0, \quad \text{if } i \neq j \tag{18}$$

$$\hat{M}_m^\dagger \hat{M}_m \geq 0, \quad \forall m. \tag{19}$$

On the other hand,

$$\hat{M}_m^n = \hat{M}_m, \quad \forall m, \tag{20}$$

with,

$$\sum_m \hat{M}_m = I \tag{21}$$

Let us work out a simple example. Assume we have a polarized photon with associated polarization orientations 'horizontal' and 'vertical'. The horizontal polarization direction is denoted by $|0\rangle$ and the vertical polarization direction is denoted by $|1\rangle$. Thus, an arbitrary initial state for our photon can be described by the quantum state $|\psi\rangle = \alpha|0\rangle + \beta|1\rangle$ (remembering Subsection 2.1), where $\alpha$ and $\beta$ are complex numbers constrained by the normalization condition $|\alpha|^2 + |\beta|^2 = 1$, and $\{|0\rangle, |1\rangle\}$ is the computational basis spanning $H^2$.

Now, we construct two measurement operators $\hat{M}_0 = |0\rangle\langle 0|$ and $\hat{M}_1 = |1\rangle\langle 1|$ and two measurement outcomes $\lambda_0, \lambda_1$. Then, the full observable used for measurement in this experiment is $\hat{M} = \lambda_0 |0\rangle\langle 0| + \lambda_1 |1\rangle\langle 1|$. According to Postulate, the probabilities of obtaining outcome $\lambda_0$ or outcome $\lambda_1$ are given by $p(\lambda_0) = |\alpha|^2$ and $p(\lambda_1) = |\beta|^2$. Corresponding post-measurement quantum states are as follows: if outcome = $\lambda_0$, then $|\psi\rangle_{pm} = |0_0\rangle_{pm} = |0\rangle$; if outcome = $\lambda_1$ then $|\psi\rangle_{pm} = |1_1\rangle_{pm} = |1\rangle$, see Eq.(29) and (30), respectively. Besides,

$$\lambda_m \geq 0, \quad \forall m, \tag{22}$$

which is equivalent to

$$|\lambda_m| = \lambda_m, \quad \forall m. \tag{23}$$

On the other hand,

$$\hat{M}^n = \lambda_0^n M_0 + \lambda_1^n M_1. \tag{24}$$

Now, considering that,

$$\hat{M}_0 |0\rangle = \begin{bmatrix} 1 & 0 \\ 0 & 0 \end{bmatrix} \begin{bmatrix} 1 \\ 0 \end{bmatrix} = \begin{bmatrix} 1 \\ 0 \end{bmatrix} = |0\rangle, \tag{25a}$$

$$\hat{M}_0 |1\rangle = \begin{bmatrix} 1 & 0 \\ 0 & 0 \end{bmatrix} \begin{bmatrix} 0 \\ 1 \end{bmatrix} = \begin{bmatrix} 0 \\ 0 \end{bmatrix} = 0 \begin{bmatrix} 0 \\ 1 \end{bmatrix} = 0|1\rangle = 0, \tag{25b}$$

$$\hat{M}_0 |\psi\rangle = \begin{bmatrix} 1 & 0 \\ 0 & 0 \end{bmatrix} \begin{bmatrix} \alpha \\ \beta \end{bmatrix} = \begin{bmatrix} \alpha \\ 0 \end{bmatrix} = \alpha \begin{bmatrix} 1 \\ 0 \end{bmatrix} = \alpha|0\rangle, \tag{25c}$$

and

$$\hat{M}_1|0\rangle = \begin{bmatrix} 0 & 0 \\ 0 & 1 \end{bmatrix}\begin{bmatrix} 1 \\ 0 \end{bmatrix} = \begin{bmatrix} 0 \\ 0 \end{bmatrix} = 0\begin{bmatrix} 1 \\ 0 \end{bmatrix} = 0|0\rangle = 0, \quad (26a)$$

$$\hat{M}_1|1\rangle = \begin{bmatrix} 0 & 0 \\ 0 & 1 \end{bmatrix}\begin{bmatrix} 0 \\ 1 \end{bmatrix} = \begin{bmatrix} 0 \\ 1 \end{bmatrix} = |1\rangle, \quad (26b)$$

$$\hat{M}_1|\psi\rangle = \begin{bmatrix} 0 & 0 \\ 0 & 1 \end{bmatrix}\begin{bmatrix} \alpha \\ \beta \end{bmatrix} = \begin{bmatrix} 0 \\ \beta \end{bmatrix} = \beta\begin{bmatrix} 0 \\ 1 \end{bmatrix} = \beta|1\rangle, \quad (26c)$$

and, if in Eq.(15) we replace $m$ by 0, and considering Eq.(25c), for the generic $|\psi\rangle$, we will have

$$|\psi_0\rangle_{pm} = \frac{\hat{M}_0|\psi\rangle}{\sqrt{\langle\psi|\hat{M}_0^\dagger\hat{M}_0|\psi\rangle}} = \frac{\alpha}{|\alpha|}|0\rangle, \quad (27)$$

and, $m$ by 1, and considering Eq.(26c), we will have

$$|\psi_1\rangle_{pm} = \frac{\hat{M}_1|\psi\rangle}{\sqrt{\langle\psi|\hat{M}_1^\dagger\hat{M}_1|\psi\rangle}} = \frac{\beta}{|\beta|}|1\rangle. \quad (28)$$

Equations (27) and (28) represent the geometric projections of $|\psi\rangle$ onto the computational basis $\{|0\rangle,|1\rangle\}$, i.e., they are the quantum measurement [1].

Finally, making the same but regarding to computational basis $\{|0\rangle,|1\rangle\}$, and using Eq.(25a) and (26c), we have

$$|0_0\rangle_{pm} = \frac{\hat{M}_0|0\rangle}{\sqrt{\langle 0|\hat{M}_0^\dagger\hat{M}_0|0\rangle}} = |0\rangle, \quad (29)$$

and

$$|1_1\rangle_{pm} = \frac{\hat{M}_1|1\rangle}{\sqrt{\langle 1|\hat{M}_1^\dagger\hat{M}_1|1\rangle}} = |1\rangle. \quad (30)$$

Equations (29) and (30) demonstrate that only CBS are not disturbed by quantum measurement [1]. Finally, using Equations (25b) and (26a), respectively, we have the cross projections, that is to say, $|1\rangle$ onto $z$ axis, and $|0\rangle$ onto $x$ and $y$ axis (see Fig.1, Bloch's sphere),

$$|1_0\rangle_{pm} = \frac{\hat{M}_0|1\rangle}{\sqrt{\langle 1|\hat{M}_0^\dagger\hat{M}_0|1\rangle}} = \frac{0}{0}, \quad (31)$$

and

$$|0_1\rangle_{pm} = \frac{\hat{M}_1|0\rangle}{\sqrt{\langle 0|\hat{M}_1^\dagger\hat{M}_1|0\rangle}} = \frac{0}{0}. \quad (32)$$

# 3 Quantum Image Processing and its implementation problems

From the above in Section 1 on this subject, it is clear that between the literature [15-44] there are serious confusions about:

- What are classical-to-quantum and quantum-to-classical interfaces?
  How to implement them physically and really? See [51-57].
- What returns a real quantum computer (non-adiabatic), in the general case? i.e., states, density matrix of such states, or what? The answer is simple, and as we can see in the last Subsection, we obtain probabilities and post measurement quantum states [58]. However, this obviousness is not clear in the papers dealing algorithms for Quantum Image Processing (QuIP) [44, 59-71]. In fact, they try classical algorithm with Dirac notation, and after such algorithm are coded in MATLAB®, in the best. Without further clarification concerning a real implementation (in laboratory) of the above algorithms, it appears that the quantum computers employed are ideal. A look at the appropriate literature helps clarify the situation [1, 72].
- What the meaning of *an appropriate internal representation of the image*? See [19-28, 31-36, 73, 74].
- What are the consequence of decoherence [57, 75-84] on the different algorithm for QuIP?
- The most of papers lacks of computational cost of treated quantum algorithm
- The most of papers did not clarify which type of quantum measurement [58, 85-93] used (weak [94-99], weak and strong, or simply a tomography [100-102]), if they used any.
- In relation to the previous point, the most of papers lacks of an estimate of the error range of the quantum measurement used [58].
- What are the consequences of quantum measurement on the outcomes?

The latter is the most serious problem, which we try to address it in detail here. On the other hand, it is noteworthy that the same problem extends to Quantum Signal Processing (QuSP) [13, 14].

*What is the quantum measurement equivalent in QuIP and QuSP?*

According to Eq.(15) of Subsection 2.3, we will have

$$|\psi_m\rangle_{pm} = \frac{\hat{M}_m|\psi\rangle}{\sqrt{\langle\psi|\hat{M}_m^\dagger \hat{M}_m|\psi\rangle}} = f_m(|\psi\rangle) \qquad (33)$$

$$= \frac{f_m(|m\rangle)}{0!}(|\psi\rangle-|m\rangle)^0 + \frac{f_m^I(|m\rangle)}{1!}(|\psi\rangle-|m\rangle)^1 + \frac{f_m^{II}(|m\rangle)}{2!}(|\psi\rangle-|m\rangle)^2 + \frac{f_m^{III}(|m\rangle)}{3!}(|\psi\rangle-|m\rangle)^3 + \ldots$$

where the second row of this equation represents its decomposition in a Taylor's series. Quickly, we can see that Eq.(31) is an odd symmetric function (see the first row of this equation), therefore the terms of even degree disappear,

$$|\psi_m\rangle_{pm} = \frac{f_m^I(|m\rangle)}{1!}(|\psi\rangle-|m\rangle)^1 + \frac{f_m^{III}(|m\rangle)}{3!}(|\psi\rangle-|m\rangle)^3 + \frac{f_m^V(|m\rangle)}{5!}(|\psi\rangle-|m\rangle)^5 + \frac{f_m^{VII}(|m\rangle)}{7!}(|\psi\rangle-|m\rangle)^7 + \ldots \quad (34)$$

Besides, at this point, we have two practical alternatives based on CBS, i.e.: $|m\rangle = |0\rangle$ (Mac Laurin's series), and $|m\rangle = |1\rangle$. In the first case, we have,

$$|\psi_0\rangle_{pm} = \frac{\hat{M}_0|\psi\rangle}{\sqrt{\langle\psi|\hat{M}_0^\dagger \hat{M}_0|\psi\rangle}} = f_0(|\psi\rangle) \qquad (35)$$

$$= \frac{f_0^I(|0\rangle)}{1!}(|\psi\rangle-|0\rangle)^1 + \frac{f_0^{III}(|0\rangle)}{3!}(|\psi\rangle-|0\rangle)^3 + \frac{f_0^V(|0\rangle)}{5!}(|\psi\rangle-|0\rangle)^5 + \frac{f_0^{VII}(|0\rangle)}{7!}(|\psi\rangle-|0\rangle)^7 + \ldots$$

with

$$f_0(|0\rangle) = |0\rangle \tag{36}$$

because $|0\rangle$ is a CBS (see Eq.29), and the Quantum Measurement does not alter the outcome. On the other hand, doing successive derivatives of the first row of Eq.(32), then, replacing $|\psi\rangle$ by $|0\rangle$, and, taking into account Equations (20) and (25a), we will have symbolically,

$$f_0^I(|\psi\rangle) = \left(\frac{\hat{M}_0}{2}\right)^1 \frac{1}{\sqrt{\langle\psi|\hat{M}_0^\dagger \hat{M}_0|\psi\rangle}} \tag{37a}$$

$$f_0^I(|0\rangle) = \frac{1}{2^1}\hat{M}_0 \tag{37b}$$

$$f_0^{II}(|\psi\rangle) = (-1)\left(\frac{\hat{M}_0}{2}\right)^2 \frac{\hat{M}_0|\psi\rangle}{\left(\sqrt{\langle\psi|\hat{M}_0^\dagger \hat{M}_0|\psi\rangle}\right)^3} \tag{38a}$$

$$f_0^{II}(|0\rangle) = \frac{(-1)}{2^2}|0\rangle \tag{38b}$$

$$f_0^{III}(|\psi\rangle) = (-1)^2\left(\frac{\hat{M}_0}{2}\right)^3 \frac{1}{\left(\sqrt{\langle\psi|\hat{M}_0^\dagger \hat{M}_0|\psi\rangle}\right)^3} \tag{39a}$$

$$f_0^{III}(|0\rangle) = \frac{(-1)^2}{2^3}\hat{M}_0 \tag{39b}$$

$$f_0^{IV}(|\psi\rangle) = (-1)^2(-3)\left(\frac{\hat{M}_0}{2}\right)^4 \frac{\hat{M}_0|\psi\rangle}{\left(\sqrt{\langle\psi|\hat{M}_0^\dagger \hat{M}_0|\psi\rangle}\right)^5} \tag{40a}$$

$$f_0^{IV}(|0\rangle) = \frac{(-1)^2(-3)}{2^4}|0\rangle \tag{40b}$$

$$f_0^V(|\psi\rangle) = (-1)^2(-3)^2\left(\frac{\hat{M}_0}{2}\right)^5 \frac{1}{\left(\sqrt{\langle\psi|\hat{M}_0^\dagger \hat{M}_0|\psi\rangle}\right)^5} \tag{41a}$$

$$f_0^V(|0\rangle) = \frac{(-1)^2(-3)^2}{2^5}\hat{M}_0 \tag{41b}$$

$$f_0^{VI}(|\psi\rangle) = (-1)^2(-3)^2(-5)\left(\frac{\hat{M}_0}{2}\right)^6 \frac{\hat{M}_0|\psi\rangle}{\left(\sqrt{\langle\psi|\hat{M}_0^\dagger \hat{M}_0|\psi\rangle}\right)^7} \tag{42a}$$

$$f_0^{VI}(|0\rangle) = \frac{(-1)^2(-3)^2(-5)}{2^6}|0\rangle \tag{42b}$$

$$f_0^{VII}(|\psi\rangle) = (-1)^2(-3)^2(-5)^2 \left(\frac{\hat{M}_0}{2}\right)^7 \frac{1}{\left(\sqrt{\langle\psi|\hat{M}_0^\dagger\hat{M}_0|\psi\rangle}\right)^7} \quad (43a)$$

$$f_0^{VII}(|0\rangle) = \frac{(-1)^2(-3)^2(-5)^2}{2^7}\hat{M}_0 \quad (43b)$$

$$f_0^{VIII}(|\psi\rangle) = (-1)^2(-3)^2(-5)^2(-7)\left(\frac{\hat{M}_0}{2}\right)^8 \frac{\hat{M}_0|\psi\rangle}{\left(\sqrt{\langle\psi|\hat{M}_0^\dagger\hat{M}_0|\psi\rangle}\right)^9} \quad (44a)$$

$$f_0^{VIII}(|0\rangle) = \frac{(-1)^2(-3)^2(-5)^2(-7)}{2^8}|0\rangle \quad (44b)$$

$$f_0^{IX}(|\psi\rangle) = (-1)^2(-3)^2(-5)^2(-7)^2\left(\frac{\hat{M}_0}{2}\right)^9 \frac{1}{\left(\sqrt{\langle\psi|\hat{M}_0^\dagger\hat{M}_0|\psi\rangle}\right)^9} \quad (45a)$$

$$f_0^{IX}(|0\rangle) = \frac{(-1)^2(-3)^2(-5)^2(-7)^2}{2^9}\hat{M}_0 \quad (45b)$$

Now, replacing Equations (37b), (39b), (41b), (43b), (45b) into the forth row of Eq.(35), we have

$$|\psi_0\rangle_{pm} = \frac{1}{2^1}\hat{M}_0\frac{(|\psi\rangle-|0\rangle)^1}{1!} + \frac{(-1)^2}{2^3}\hat{M}_0\frac{(|\psi\rangle-|0\rangle)^3}{3!} + \frac{(-1)^2(-3)^2}{2^5}\hat{M}_0\frac{(|\psi\rangle-|0\rangle)^5}{5!} + \\ \frac{(-1)^2(-3)^2(-5)^2}{2^7}\hat{M}_0\frac{(|\psi\rangle-|0\rangle)^7}{7!} + \frac{(-1)^2(-3)^2(-5)^2(-7)^2}{2^9}\hat{M}_0\frac{(|\psi\rangle-|0\rangle)^9}{9!} + \ldots \quad (46)$$

Obviously, if $|\psi\rangle = |0\rangle$ (outcome = $\lambda_0$), then $|0_0\rangle_{pm} = |0\rangle$ (satisfying Eq.29), i.e., it is a CBS case where Quantum Measurement does not alter the outcome. On the other hand, if $|m\rangle = |1\rangle$ and considering Eq.(33) is an odd symmetric function (this is identical to the previous case), thus, it will be

$$|\psi_1\rangle_{pm} = \frac{\hat{M}_1|\psi\rangle}{\sqrt{\langle\psi|\hat{M}_1^\dagger\hat{M}_1|\psi\rangle}} = f_1(|\psi\rangle) \\ = \frac{f_1^I(|1\rangle)}{1!}(|\psi\rangle-|1\rangle)^1 + \frac{f_1^{III}(|1\rangle)}{3!}(|\psi\rangle-|1\rangle)^3 + \frac{f_1^V(|1\rangle)}{5!}(|\psi\rangle-|1\rangle)^5 + \ldots \quad (47)$$

with

$$f_1(|1\rangle) = |1\rangle \quad (48)$$

because $|1\rangle$ is a CBS (see Eq.30), and the Quantum Measurement does not alter the outcome.

On the other hand, doing successive derivatives of the first row of Eq.(47), then, replacing $|\psi\rangle$ by $|1\rangle$, and, taking into account Equations (20) and (26b), and with similar considerations with respect to the previous case, we will have symbolically and finally,

$$|\psi_1\rangle_{pm} = \frac{1}{2^1}\hat{M}_1\frac{(|\psi\rangle-|1\rangle)^1}{1!} + \frac{(-1)^2}{2^3}\hat{M}_1\frac{(|\psi\rangle-|1\rangle)^3}{3!} + \frac{(-1)^2(-3)^2}{2^5}\hat{M}_1\frac{(|\psi\rangle-|1\rangle)^5}{5!} + \qquad (49)$$
$$\frac{(-1)^2(-3)^2(-5)^2}{2^7}\hat{M}_1\frac{(|\psi\rangle-|1\rangle)^7}{7!} + \frac{(-1)^2(-3)^2(-5)^2(-7)^2}{2^9}\hat{M}_1\frac{(|\psi\rangle-|1\rangle)^9}{9!} + \cdots$$

Obviously, if $|\psi\rangle = |1\rangle$ (outcome = $\lambda_1$), then $|1_1\rangle_{pm} = |1\rangle$ (satisfying Eq.30), i.e., it another CBS case where, here too, Quantum Measurement does not alter the outcome.

Now, regrouping terms of Eq.(46), and remember Equations (25c), we have

$$|\psi_0\rangle_{pm} = \frac{\hat{M}_0}{2}|\psi\rangle + n_{0,QM} \qquad (50)$$
$$= \frac{\alpha}{2}|0\rangle + n_{0,QM}$$

where, $n_{0,QM}$ is a residue owing to the higher degree components. In fact, both Equation 46 and the 49, the terms of degree greater than 1 result from the power (with exponent $\geq 3$) for components with module $\leq 1$, and divided by the corresponding factorials and powers of 2, therefore, they have a rather attenuated impact on the term of degree equal to 1. Besides, from Stochastic Processes [55] we know that $n_{0,QM}$ is a noise due to the quantum measurement, in fact, its subscript means just that, while, $\hat{M}_0/2$ represents the measurement matrix. That is to say, the very act of measuring introduces noise, since, it is a disturbance. On the other hand, from Eq.(27) we can estimate the mean value (considering the errors of measurement in this projection) of such residue, that is to say,

$$n_{0,QM} = \frac{\alpha(2-|\alpha|)}{2|\alpha|}|0\rangle \qquad (51)$$

The precision involved in this estimation depends on the method of employed quantum measurement [53]. In the same way, if we have grouped terms of Eq.(49), and remember Equations (26b), we have

$$|\psi_1\rangle_{pm} = \frac{\hat{M}_1}{2}|\psi\rangle + n_{1,QM} \qquad (52)$$
$$= \frac{\beta}{2}|1\rangle + n_{1,QM}$$

and through a reasoning identical to the previous case, and taking in account Eq.(28), we come to an estimation of the mean value of $n_{1,QM}$ with similar considerations to those established for the previous case, regarding to an analysis with focusing in Stochastic Processes. Then,

$$n_{1,QM} = \frac{\beta(2-|\beta|)}{2|\beta|}|1\rangle \qquad (53)$$

Equations (50) and (52) tell us that when we measure $|\psi\rangle$ (in each of its projections, see Equations 27 and 28), we are introducing a noise measurement, which is inherent to the type of measurement, however, always exists and it cannot be overlooked, and this noise will disappear only when we measure CBS $\{|0\rangle,|1\rangle\}$. In other words, when we measure $|\psi\rangle$ (generic case), we do not have the value of each projection direct, clean and clearly, i.e., the values obtained are corrupted by noise.

In case of generic qubits, i.e., Eq. (1), the question is: what is possible without disturbing partially known quantum states? The answer is: nothing [104, 105]. Besides, optimal, reliable estimation of quantum states is only theoretical [106].

Now, if we express Eq.(6) in a more general discrete form, we will obtain

$$\left|\psi_{t_2}\right\rangle = U_{t_1,t_2}\left|\psi_{t_1}\right\rangle \qquad (54)$$

being $U_{t_1,t_2}$ the quantum algorithm for QuIP. We can see this for each pixel in Fig.4.

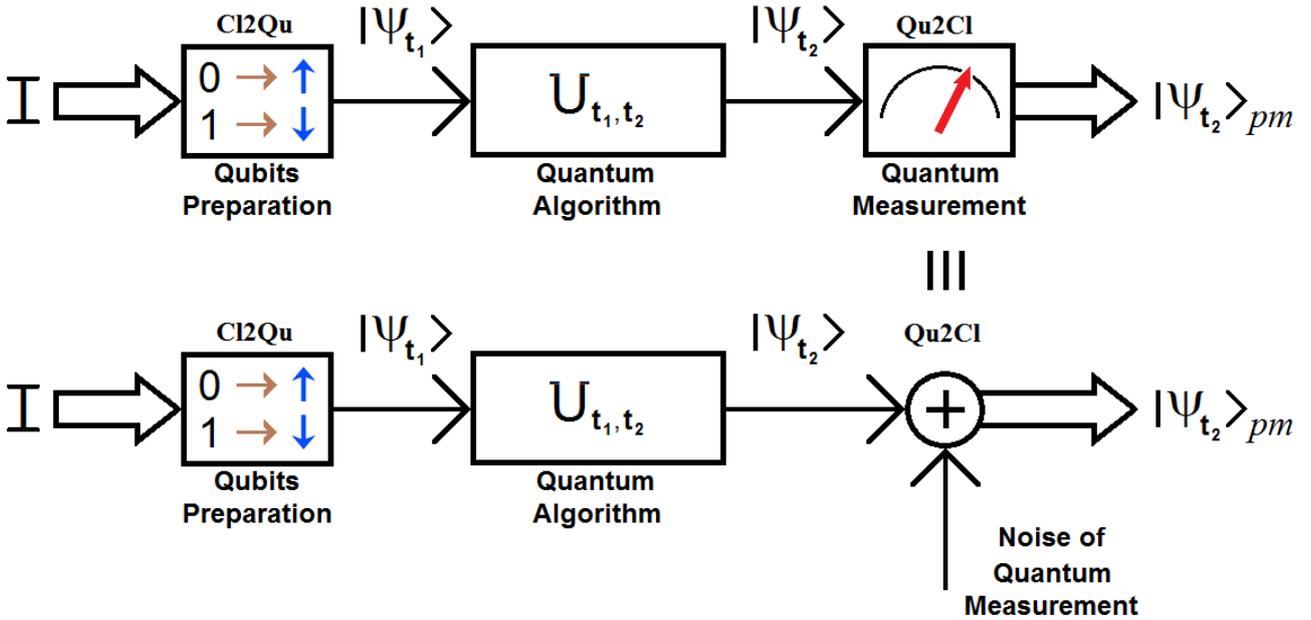

**Fig. 4** Equivalence between quantum measurement and measurement noise.

In fact, in Fig.4 we can see the classical image $I$, and the post-measurement outcome $\left|\psi_{t_2}\right\rangle_{pm}$, where Cl2Qu means *part of classical-to-quantum interface*, and Qu2Cl means *part of quantum-to-classical interface*. Figure 4 shows clearly that represented by the Eq.(29), that is to say, quantum measurement becoming a noise. Besides, $\left|\psi_{t_2}\right\rangle_{pm}$ is the recovered image after the quantum image processing, i.e., it is a classical image. Now, let's look at a couple of very conspicuous cases.

***Case #1 - the quantum algorithm does nothing***: this case can be seen on Fig.5. Here $\left|\psi_{t_2}\right\rangle = \left|\psi_{t_1}\right\rangle$, however, quantum measurement introduces noise anyway, then, the first conclusion that we arrived is:

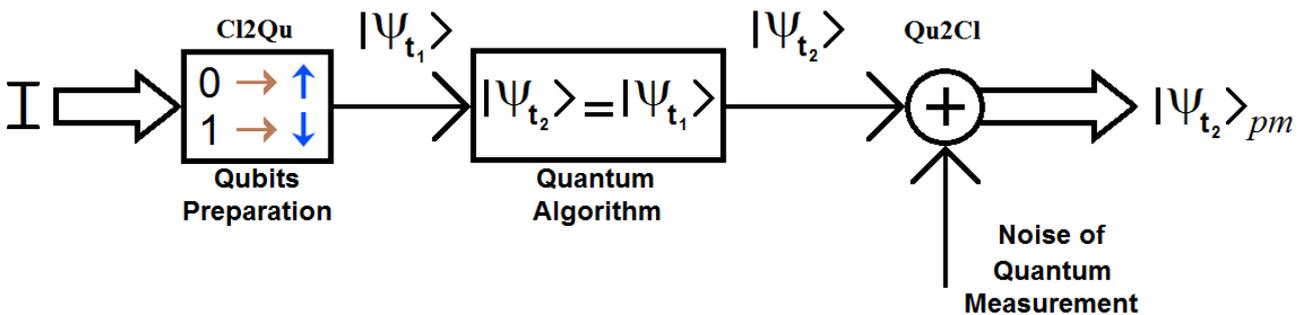

**Fig. 5** The quantum algorithm does nothing.

how is possible to store images in a quantum computer? It is obvious that trying to recover these images we introduce noise to them. Then, we need a classical filter to image denoising at the output, where $\hat{I}$ means estimated $I$, see Fig.6. Obviously, it is absurd. However, there is much literature on this topic [18, 29, 30, 107-111], which occurs on that mistake. How is possible?

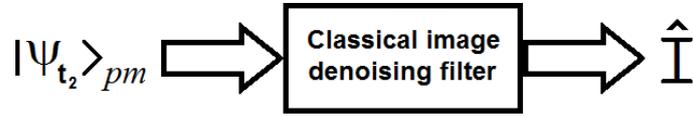

**Fig. 6** Classical image denoising filtering after quantum measurement. In reality, after quantum-to-classical interface.

***Case #2 - the quantum algorithm filters images***: this case can be seen on Fig.7. Here $|\psi_{t_2}\rangle = U_{t_1,t_2}|\psi_{t_1}\rangle$, where $U_{t_1,t_2}$ represents the quantum algorithm for image denoising [112, 113]. In Fig.7 we can see that at the entrance a noise adds to the image (image and noise are classic). This noise is known as noise of state, and this can be generated in two ways:
  a) It brings the classic original image that must be filtered
  b) It is the result of a suboptimal preparation of the qubits

Remembering Eq.(52) and considering that $t_1 = t$, and $t_2 = t_1 + \Delta t$, we can represent the intervention of this noise as follows:

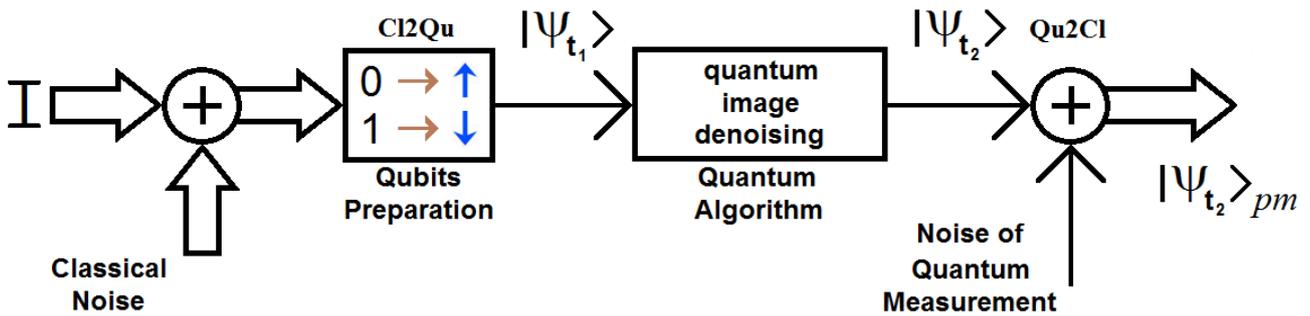

**Fig. 7** Quantum image denoising algorithm for the filtering of classical noise.

$$|\psi_{t+\Delta t}\rangle = U_{t,t+\Delta t}|\psi_t\rangle + n_{state} \tag{55}$$

Equation (53) is known as equation of state, which, together with the equations (48) and (50) constitute the general model of any linear (or linearizable) system affected simultaneously by state and measurement noise at the same time, where $n_{state}$ is the noise of state, while $U_{t,t+\Delta t}$ is the matrix of state [114-121].

Under these conditions, we could only estimate optimally the state if we knew the statistics of both noise. Obviously, this is not the case in quantum measurement [106].

Therefore, in this case we also need a filter to the output (i.e., in the classic environment, and as in the previous case, see Fig.6), that is to say, re-filter but in the classic world, where it was the original classic noise. Then, why do we use a quantum computer? Ergo, why do we use Quantum Image Processing? Absurd. Summing-up, we cannot dispense with the external filter in any case.

The quantum measurement can put more noise than the original classic noise, at most of the same order, or worse yet, we don't know.

The same is true in Quantum Image Segmentation [122, 123], and Quantum Image Encryption [37-43, 124-131], and any another intervention of Quantum Image Processing. In fact, the foregoing is extended to any quantum algorithm [1], including Quantum Signal Processing [13, 14]. Although it is incredible, no paper on quantum algorithms (with generic qubits) performs an exhaustive analysis of the robustness of the proposed algorithm.

On the other hand, returning to Section I, if we have and uncertainty in the quantum measurement equal to $\Delta_{QM}$ (e.g., 6 %), then, we will have a spread of $2 \times \Delta_{QM}$ (e.g., 12 %). It is obvious that this has a nasty impact on both the visual intelligibility of the image as well as in the quality of the definition of their edges, in its texture, etc. In after words, this is unacceptable from the point of view of Digital Image Processing [45-48].

All this can be seen clearly in Figures 8 (Agus in Miami), 9 (Angelina) and 10 (Lena). The three figures show original image, *the resultant* noisy image by fault of Quantum Measurement, *a* mesh with the noise of Quantum Measurement, and, the noise by Quantum Measurement in a colormap scale.

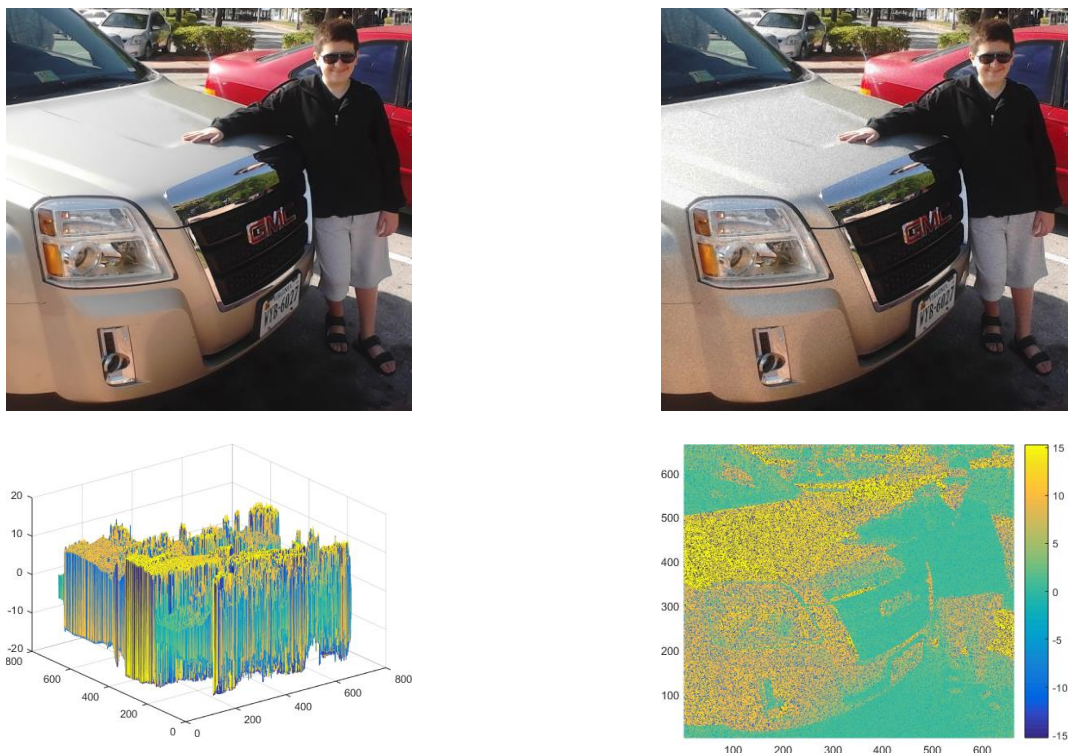

**Fig. 8** Agus in Miami: *top-left* is original image, *top-right* is noisy image because of Quantum Measurement, *down-left* is a mesh with the noise of Quantum Measurement, and, *down-right* is the noise by Quantum Measurement in a colormap scale.

This obliges us to filter in the classical world (outside of the quantum computer) and then we must to restore the edges through masks of enhancement [45-48]. That is to say, we have more activity outside quantum computer than inside quantum computer. Is this logical? Is this practical? Justifies this the use of a quantum computer to process images? All these problems are a direct result of underestimating the quantum measurement. If the problem of quantum measurement would be solved, in such a way that it could be overlooked (or at best underestimate) as indeed happens in "whole" literature on Quantum Image Processing [15-44,59-71,73,74,107-131], only then we could replace the notation of the classic algorithms of Digital Image Processing with Dirac's notation, and ready! But this is not the case. The only act of measure collapses the wave function since the measurement is a disturbance that manifests as an equivalent noise of measurement. The impact of such noise can be saw in a quantitative way in the next Sections which include a set of conspicuous simulations.

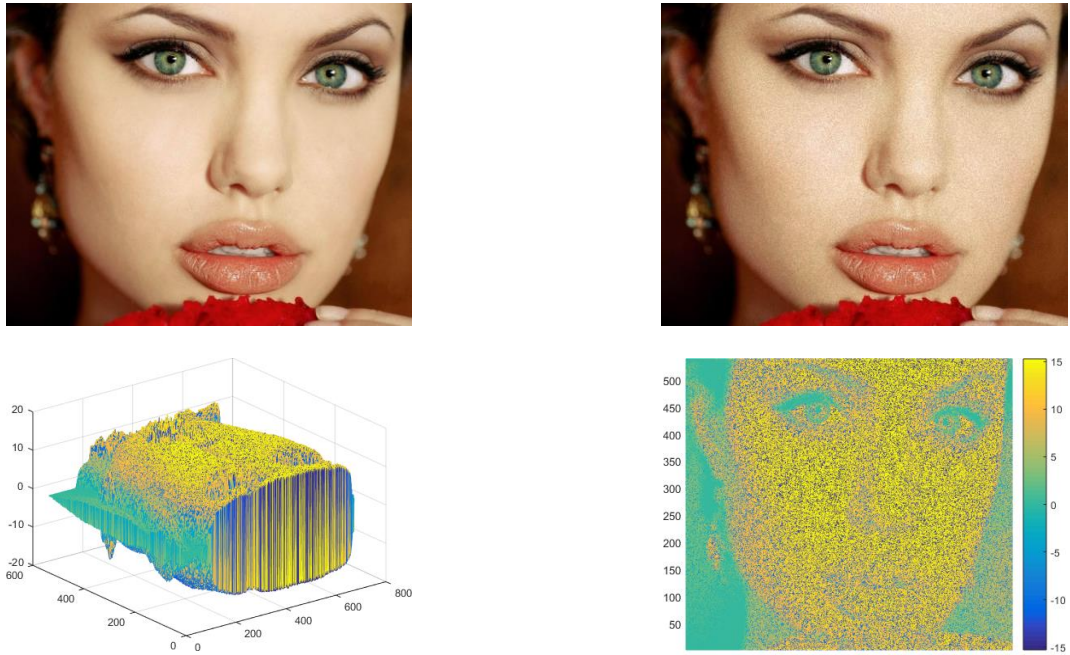

**Fig. 9** Angelina: *top-left* is original image, *top-right* is noisy image because of Quantum Measurement, *down-left* is a mesh with the noise of Quantum Measurement, and, *down-right* is the noise by Quantum Measurement in a colormap scale.

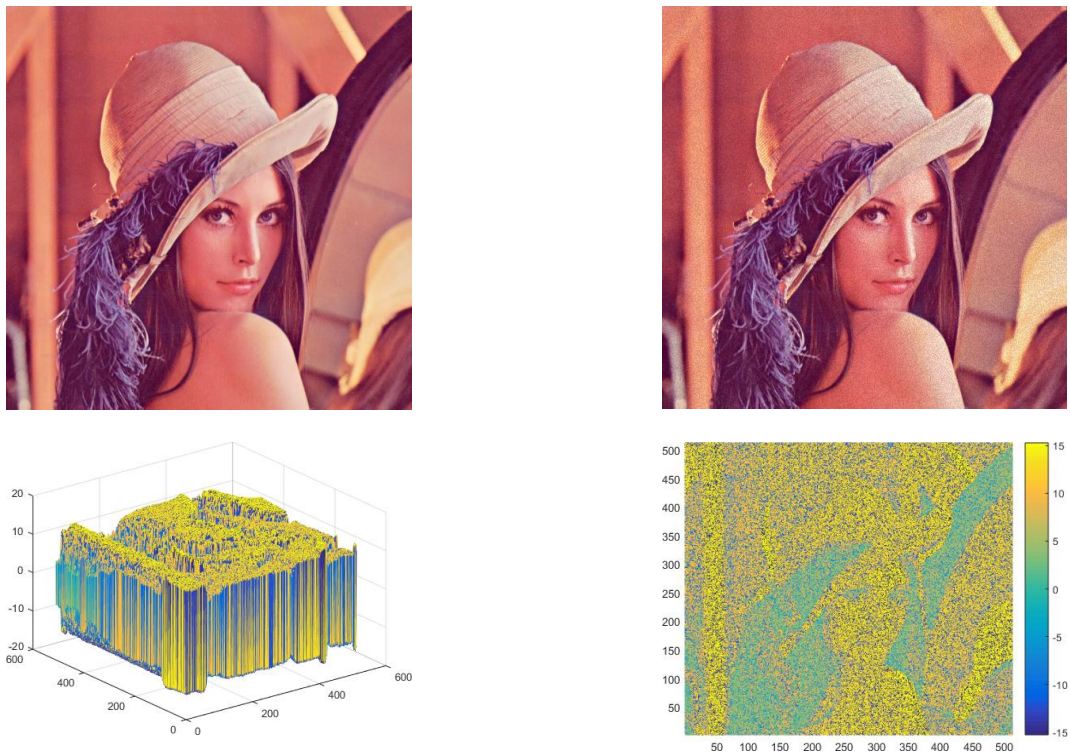

**Fig. 10** Lena: *top-left* is original image, *top-right* is noisy image because of Quantum Measurement, *down-left* is a mesh with the noise of Quantum Measurement, and, *down-right* is the noise by Quantum Measurement in a colormap scale.

A direct consequence of this is that we cannot store infinite information in the infinite decimals of both projections of a generic qubit. See Eq.(3). In reality, we could. What we cannot do is to know this information because the quantum measurement would alter the stored values.

# 4 Prolegomena to experiments and simulations

In this section, we present a series of tools which are necessaries for an important set of comparative simulations. We consider such simulations are essential for the purpose of establishing the central idea of this paper. The mentioned tools are:
- Quantum Boolean Image Processing (QuBoIP) [50]
- Flexible Representation of Quantum Images (FRQI) [22]
- Novel Enhanced Quantum Representation (NEQR) [32]
- Quantum State Tomography (QuST) [100-102,132]

In the Section 5 (which will be fully dedicated to simulations), the first three tools will be compared between them, for which we take as a reference the values estimated in QuST, i.e., we will calculate the measurement errors of QuBoIP, FRQI and NEQR against such estimated values. Further, we are going to do a comparative analysis of different aspects related to performance of mentioned techniques.

Other techniques for the internal representation of the image are Qubit Lattice [19], Entangled Image [20], Real Ket [21], Multi-channel representation of quantum images (MCQI) [63], Normal arbitrary superposition state (NASS) [30], Normal arbitrary quantum superposition state (NAQSS) [108], Geometric Quantum Image Representation (GQIR) [69], Quantum representation for color digital images (QUALPI) [34], QSMC represents M colors and QSNC represents the coordinates of N pixels in an image (QSMC&QSNC) [29], Color quantum image based on phase transform (CQIPT) [73], Improvement of FRQI (IFRQI) [133], Geometric transformations on quantum images (GTQI) [134], Improved novel enhanced quantum representation (INEQR) [135], and Multi-channel representation for quantum images (MCRQI) [136]. This techniques also have been tested and have not given better results than QuBoIP. Therefore, in Section 5 we will present a set of interesting and comparative simulation based on FRQI, NEQR, and QUBoIP, whereby, the central idea of this work is set sufficiently highlighted.

***Quantum Boolean Image Processing (QuBoIP):*** this technology is presented as a viable alternative to the problem of practical implementation of algorithms for quantum image processing (QuIP) [50]. The algorithms grouped under this technology does not suffer the ravages of the quantum measurement. However, QuBoIP and QuIP share a common enemy: the decoherence [57, 75-84]. Of all forms, and as it is logical to assume, decoherence affects a lot more to the QuIP that the QuBoIP, given that QuBoIP works with CBS, it is more easy to operate with threshold logic, until the ultimate collapse. On the other hand, and as we can see in previous sections, there is a direct and automatic correspondence between $\{0,1\}$ and $\{|0\rangle,|1\rangle\}$. Such correspondence (and in that order) will be the classical-to-quantum interface. In the same way, but in reverse order, there is a direct and automatic correspondence between $\{|0\rangle,|1\rangle\}$ and $\{0,1\}$. In this correspondence, but in that order, we know it as a quantum-to-classical interface. Unlike of QuIP, in QuBoIP the measurement is not a problem, since it does not alter the outcome measure. Therefore, in QuBoIP is not necessary a filter (and more) after quantum measurement [50], as in QuIP.

In Table I we can see these statements, where left column represents the state before measurement, while right column represents the state after that, for CBS and generic state (see Subsection 2.1), where $\hat{M}_m$ is the *measurement operator* (see Eq.14), and $|\psi_m\rangle_{pm}$ is the *postmeasurement quantum state* in its generic form [1].

TABLE I
MEASUREMENT OUTCOME WITH CBS AND GENERIC STATE.

| Before quantum measurement | After quantum measurement |
|---|---|
| $|0\rangle$ | $|0\rangle$ |
| $|1\rangle$ | $|1\rangle$ |
| $|\psi\rangle$ | $|\psi_m\rangle_{pm} = \dfrac{\hat{M}_m|\psi\rangle}{\sqrt{\langle\psi|\hat{M}_m^\dagger \hat{M}_m|\psi\rangle}}$ |

That is to say, here too, it is only necessary to consider the poles of Bloch's sphere when we work with CBS (see poles of Figures 1 and 2), i.e., when we working with one qubit only, which is all that is used in this technology. Besides, while in the case of CBS before and after quantum measurements are matching situations, for generic qubits (third row of Table I) the case is completely different [1, 58].

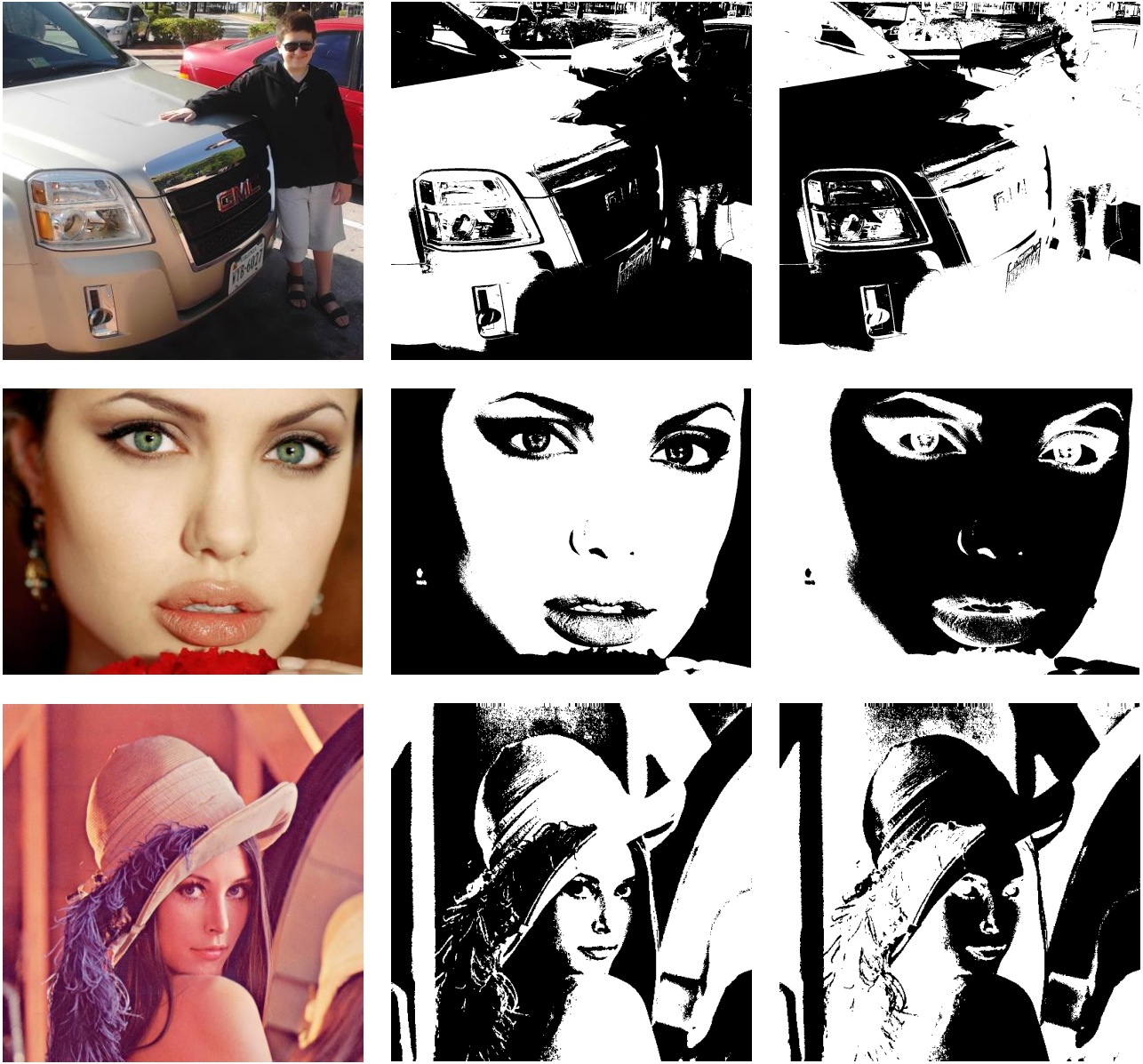

**Fig. 11** Quantum entanglement in QuBoIP between second and third columns of each case: first row for *Agus in Miami*, second row for *Angelina*, and third row for *Lena*. The first column is composed by the original images, while, the second column is constituted by the MSB of gray version of each image. Then, the second and third column are entangled between them.

Table I is the cornerstone of this methodology called quantum Boolean image processing in general, and quantum Boolean image denoising in particular. It will allow us (among others):

a) to build more robust interfaces with respect to quantum measurement noise, decoherence, etc.,

b) to ignore the problem of quantum measurement [1, 58], which was above mentioned,

c) a design more close to the hardware,

d) a lower computational and memory cost [50], working only with the *most significant bit* (MSB) and its quantum counterpart, in simulations of quantum computing on classical systems like GPGPU [137], FPGA [138], given that in a real and future quantum computer becomes irrelevant an evaluation and consideration regarding to computational cost, because, this technology promises to introduce a huge quantity of quantum processors (molecules or particles) in an incredibly small nanometric space, allowing us face great computational challenges, which are unacceptable today for the classic computers, such as the problems of the type NP complete.
e) to export this criterion beyond the QuIP, e.g., to Quantum Signal Processing [13, 14], and
f) among others.

Working with CBS frees us from the influence that superposition have normally when we work with generic cubits. In other words, we are in a very special borderline case, i.e., it is similar to classical Boolean [50]. In fact, there is no difference between quantum Boolean and pure Boolean (classical) regarding to their results in a filtering context. The quantum Boolean version has the same computational cost of classical Boolean version. Besides, for this algorithm, we will seek that $|\psi\rangle$ be CBS (all the time), because it's easier its state treatment in presence of quantum decoherence. On the other hand, the quantum decoherence time depends on the implementation technology of the quantum computer and not the quantum algorithm itself [1]. However, it is easier a collapse of the wave function with generic qubits that with CBS, the state is more stable in a quantum Boolean context than in a generic quantum context, in special after measurement of the state.

Besides, QuBoIP allow us work in a more practical way with the entanglement. Figure 11 show us quantum entanglement in QuBoIP between second and third columns of each case: first row for *Agus in Miami*, second row for *Angelina*, and third row for *Lena*. The first column is composed by the original images, while, the second column is constituted by the MSB of gray version of each image. Then, the second and third column are entangled between them. This is easily verifiable in the laboratory for each most significant qubit of the image.

For everything that we said, we can see that QuIP can be implemented in a practical way through techniques QuBoIP [50].

***Flexible Representation of Quantum Images (FRQI):*** As we can see in [32], every pixel is mapped into a basis state of a four-dimensional qubit sequence. FRQI was proposed by Le [22]. This quantum image-representation model can be expressed as in (56) for a $2^n \times 2^n$ image. In FRQI, the position information of every pixel is stored in a basis state of a two-dimensional qubit sequence, and the gray-scale information is stored as the probability amplitude of a single qubit which is entangled with the qubit sequence. Figure 12 shows a $4 \times 4$ FRQI quantum image.

$$|I\rangle = \frac{1}{2^n} \sum_{Y=0}^{2^n-1} \sum_{X=0}^{2^n-1} \left( cos\,\theta_{YX} |0\rangle + sin\,\theta_{YX} |1\rangle \right) |YX\rangle \tag{56}$$

Qubit Lattice and Entangled Image are similar to classical digital image processing, and therefore it is easy to find the quantum equivalents for classical image-processing operations, yet without much performance improvement. The Real Ket and FRQI models use the superposition of a qubit sequence to store images, and therefore they can process information for all pixels simultaneously. Compared with Real Ket, FRQI maintains a 2D pixel matrix, and therefore it is the most suitable and flexible model for designing quantum image-processing algorithms among all the existing quantum image representations.

Many studies on quantum image processing have been carried out based on FRQI [32,64] discussed simple geometric and color operations based on FRQI and proved the great performance improvement of these quantum operations compared with classical image operations expanded FRQI using three color qubits to store the full-color information of an RGBα image designed quantum watermarking algorithms based on FRQI, obtaining good performance. In general, all previous studies based on FRQI have focused on certain simple operations and algorithms in digital image processing. The main reason for this is that the FRQI model has some drawbacks:

1. The time complexity of quantum image preparation for FRQI is too high. For a $2^n \times 2^n$ image, the procedure costs $O(2^{4n})$, which is quadratic in the image size.
2. To simplify preparation of an FRQI quantum image, the Boolean expression minimization method is used to perform image compression. However, it has been found that the FRQI compression ratio depends on the gray-scale distribution of the digital image; especially when the image is highly disordered, the FRQI compression ratio is very low, even approaching zero.
3. Because FRQI stores the gray-scale information of the image pixels as the probability amplitude of a single qubit, it is impossible to obtain accurate probability amplitude for this qubit through finite quantum measurements. In other words, accurate image retrieval is an impossible task for the FRQI model.
4. FRQI stores the digital image into the superposition of a qubit sequence according to the position information of different pixels. Therefore, all operations which are suitable for FRQI should perform the same operations for pixels at different positions. Other operations cannot be accommodated.

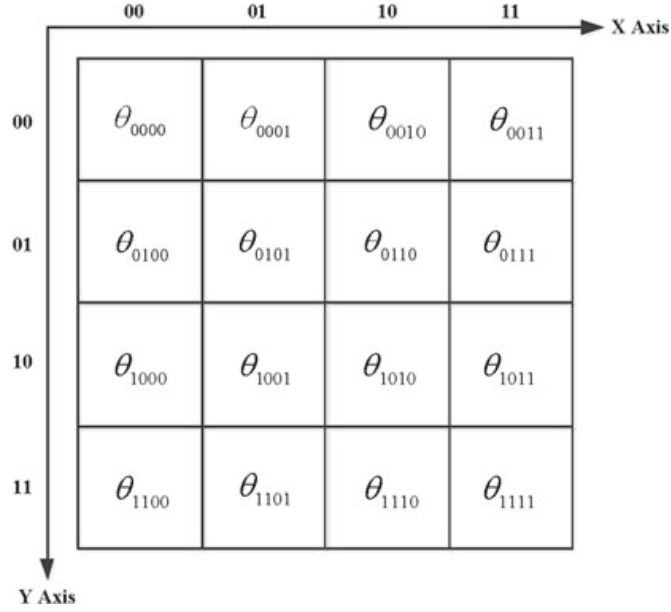

**Fig. 12** A 4×4 FRQI quantum image. $\theta_{YX}$ is the phase of the color qubit which denotes the gray-scale value of pixel *(Y, X)*.

Besides, from [64] we can see an example of a 2x2 FRQI quantum image with its corresponding state is presented in Fig.13.

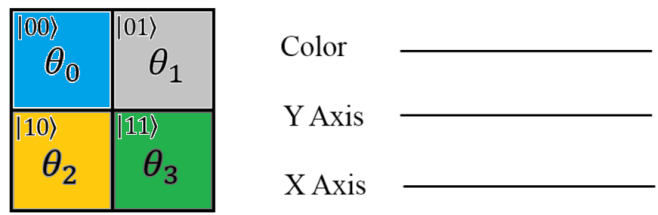

$$|I\rangle = \frac{1}{2}[(\cos\theta_0|0\rangle + \sin\theta_0|1\rangle) \otimes |00\rangle + (\cos\theta_1|0\rangle + \sin\theta_1|1\rangle) \otimes |01\rangle$$
$$+(\cos\theta_2|0\rangle + \sin\theta_2|1\rangle) \otimes |10\rangle + (\cos\theta_3|0\rangle + \sin\theta_3|1\rangle) \otimes |11\rangle]$$

**Fig. 13** A simple FRQI image and its quantum state.

To effectively capture all the information about the colour and position of a $2^n \times 2^n$-sized FRQI image, a total of *2n+1* qubits are required as shown in the generalised circuit of the FRQI representation in Fig.14.

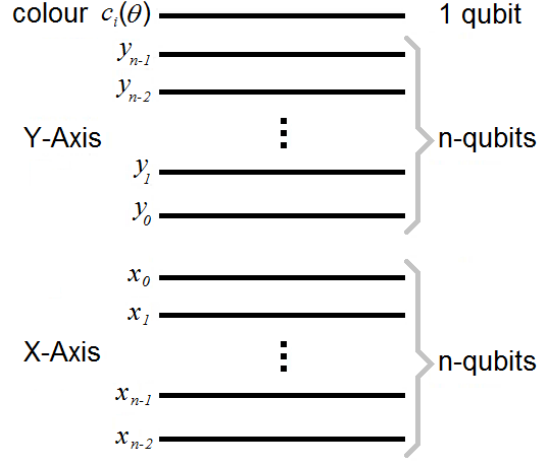

**Fig. 14** Generalised circuit showing how information in an FRQI quantum image state is encoded.

These disadvantages of FRQI have become limitations on the exploration of complex quantum image-processing algorithms. Hence, it is of significant importance to improve existing quantum image models and to find better ways to represent quantum images for the future development of quantum image processing.

*Novel Enhanced Quantum Representation (NEQR):* As we can see in [32], the advantages of FRQI result from using the superposition of a qubit sequence to store the position information of all the pixels, so that all can be operated on simultaneously. However, the main reason for the drawbacks of this model is that FRQI uses only a single qubit to store the gray-scale information for each pixel. To improve FRQI, the NEQR model uses two entangled qubit sequences to store the gray-scale and position information, and stores the whole image in the superposition of the two qubit sequences.

Suppose the gray range of image is $2^q$, binary sequence $C_{YX}^0 C_{YX}^1 \ldots C_{YX}^{q-2} C_{YX}^{q-1}$ encodes the gray-scale value $f(Y,X)$ of the corresponding pixel $(Y,X)$ as in (57):

$$f(Y,X) = C_{YX}^0 C_{YX}^1 \ldots C_{YX}^{q-2} C_{YX}^{q-1}, C_{YX}^k \in [0,1], \quad f(Y,X) \in [0, 2^q - 1] \tag{57}$$

The representative expression of a quantum image for a $2^n \times 2^n$ image can be written as in (58):

$$|I\rangle = \frac{1}{2^n} \sum_{Y=0}^{2^n-1} \sum_{X=0}^{2^n-1} |f(Y,X)\rangle |YX\rangle = \frac{1}{2^n} \sum_{Y=0}^{2^n-1} \sum_{X=0}^{2^n-1} \bigotimes_{i=0}^{q-1} |C_{YX}^i\rangle |YX\rangle \tag{58}$$

Figure 15 shows a $4 \times 4$ NEQR image example. Compared with the example of FRQI in Fig. 12, the obvious difference is that NEQR utilizes the basis state of qubit sequence to represent the gray scale of pixels instead of probability amplitude of a single qubit in FRQI. Different gray scales can be distinguished in NEQR because different basis states of qubit sequence are orthogonal. All the drastic improvements of NEQR in the following discussion are depended on this peculiar change.

Figure 16 illustrates a $2 \times 2$ gray-scale image and its representative expression in NEQR. In this figure, because the gray scale ranges between 0 and 255, eight qubits are needed in NEQR to store the gray-scale information for the pixels. Therefore, NEQR needs *q+2n* qubits to represent a $2^n \times 2^n$ image with gray range $2^q$.

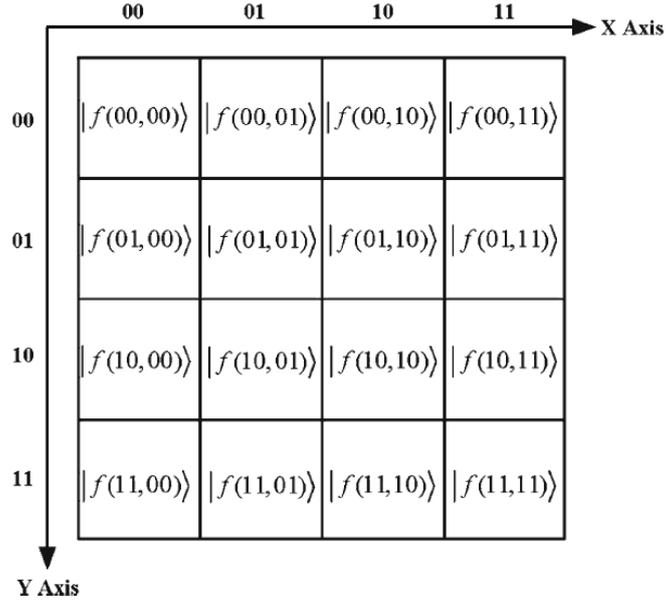

**Fig. 15** A 4×4 NEQR quantum image. $f(Y, X)$ denotes the gray-scale value of pixel $(Y, X)$, which is stored as the basis state $|f(Y,X)\rangle$ of a qubit sequence.

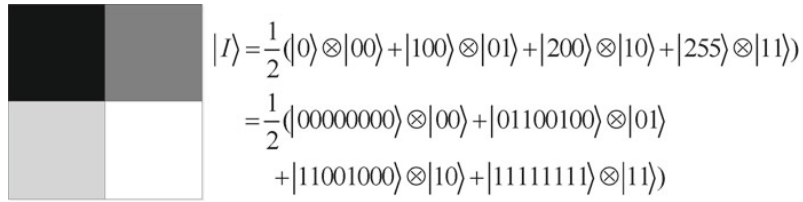

**Fig. 16** A 2×2 example image and its representative expression in NEQR.

***Quantum State Tomography (QuST):*** As we can see in [100-102,132], this technique is very important in the context of Quantum Measurement problem for Quantum Information Processing *in general*, and Quantum Image Processing *in particular*. Besides, from [139] we know that for pure states like Equations 1, 2 and 3, the density matrix is:

$$\rho = |\psi\rangle\langle\psi| \qquad (59)$$

However, it is possible to define a statistical mixture of pure states (called a mixed state) using the density operator, which is defined as:

$$\rho = \sum_i P_i |\psi_i\rangle\langle\psi_i| \qquad (60)$$

where $P_i$ is the probability of obtaining state $|\psi_i\rangle$. The sum can in general extend arbitrarily far, but any single qubit density operator can be diagonalized (in the linear algebraic sense) and represented as the probabilistic sum of just two states. For a pure state, $P_1 = 0$ and $P_0 = 1$. The density operator is Hermitian and must only have positive eigenvalues since these eigenvalues correspond to the probabilities for obtaining $\psi_i$ in the eigenvector basis. In addition, the probabilities $P_i$ must sum to one. Since the states $\psi_i$ are normalized, this condition can be concisely written as $\text{Tr}(\rho) = 1$. The purity of the state is a useful quantity that can be defined as

$$P(\rho) = \sum_i P_i^2 \leq 1 \tag{61}$$

We will see later that any density operator can be written as a point within the Bloch sphere, with the pure states lying on the surface.

The goal of state tomography is to produce an estimate of an unknown quantum state. Typically, in a real-world scenario an experimentalist is given a certain number of copies of an unknown state (or equivalently, a limited time in which she must perform measurements). The experimentalist then performs measurements on her copies of the unknown state and from these measurements she produces an estimate. This situation is depicted in Fig.17. The number of measurements the experimentalist must perform is dependent on the size of the quantum state being measured. In [139] they stated that for an n-qubit state there are $2^{2n}-1$ free parameters that describe the state. For example, looking at Eq.(62),

$$\rho = \begin{pmatrix} 1 + \langle \sigma_z \rangle & \langle \sigma_x \rangle + i \langle \sigma_y \rangle \\ \langle \sigma_x \rangle - i \langle \sigma_y \rangle & 1 - \langle \sigma_z \rangle \end{pmatrix} \tag{62}$$

it is obvious that in order to determine a single qubit state one must measure the expectation values of the three Pauli operators [139]

$$\sigma_x = \begin{pmatrix} 0 & 1 \\ 1 & 0 \end{pmatrix}, \quad \sigma_y = \begin{pmatrix} 0 & -i \\ i & 0 \end{pmatrix}, \quad \sigma_x = \begin{pmatrix} 1 & 0 \\ 0 & -1 \end{pmatrix}. \tag{63}$$

while, the purity of the state is simply:

$$P(\rho) = \frac{1}{2}\left(1 + \langle \sigma_x \rangle^2 + \langle \sigma_y \rangle^2 + \langle \sigma_z \rangle^2\right) = \frac{1}{2}(1 + r^2) \tag{64}$$

where $r$ is the radius of the Bloch vector. In our geometric picture of the single qubit, this corresponds to measuring the Bloch vector's projection onto three orthogonal axes. It is often the case that a parametrization that yields parameters which are directly measurable is not possible. In this case, each measurement made will yield a linear combination of parameters. After a sufficient number of measurements is made, this system of equations - which is linear in the parameters - can be inverted. This technique, aptly named linear inversion, is a form of tomography.

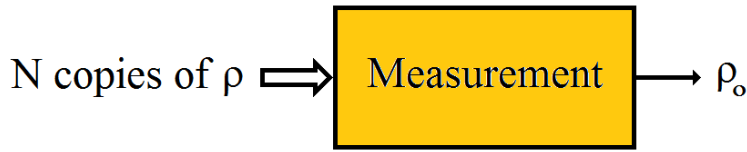

**Fig. 17** A depiction of state tomography. N copies of an unknown state are measured, and from the outcomes of these measurements an estimate $\rho_0$ for the unknown density matrix $\rho$ is obtained.

We begin by writing a general density matrix in terms of a minimal set of unknown parameters. For example, this can be done for a two qubits by writing:

$$\rho = \begin{pmatrix} t_1 & t_5 + it_{11} & t_6 + it_{12} & t_7 + it_{13} \\ t_5 - it_{11} & t_2 & t_8 + it_{14} & t_9 + it_{15} \\ t_6 - it_{12} & t_8 - it_{14} & t_3 & t_{10} + it_{16} \\ t_7 - it_{13} & t_9 - it_{15} & t_{10} - it_{16} & t_4 \end{pmatrix} \tag{65}$$

In experiments with photons, usually a rate of detected photons is measured. I will show in [139] how these photon rates are converted to probabilities in the lab, but for now we will assume that some absolute rate $N$ is known, so the rates $\eta$ are converted to probabilities (technically these are called "observed frequencies" since one cannot 'observe' a probability) as $\mu_i = \frac{\eta_i}{N}$. Then for each measurement setting, we have:

$$\mu_i = \frac{\eta_i}{N} = Tr(\hat{O}_i \rho) = \sum_j M_{ij} t_j \qquad (66)$$

where the measurement operator $\hat{O}$ is said to be a projector onto the state $|\phi\rangle$ if it takes the form:

$$\hat{O} = |\phi\rangle\langle\phi| \qquad (67)$$

being $Tr(\bullet)$ the trace of $(\bullet)$. On the other hand, Eq.(66) along with the equation $t_1 + t_2 + t_3 + t_4 = 1$ (for normalization) can be written as

$$\mu = M t \qquad (68)$$

One need only invert the matrix $M$ to solve this system of linear equations. There is a problem with linear inversion, however. The form of Eq.(65) does not constrain $\rho$ sufficiently. In an experiment, each of the expectation values can only be estimated with a finite number of copies of the unknown state. The estimates of the expectation values are all constrained to be within the eigenvalue spectrum of the operators that they represent (there is no way to obtain values outside of the eigenvalue spectrum), but the sum of the squares of the expectation values are not constrained at all. As we can see in [139], the sum of the squares of the expectation values of the Pauli operators is the radius of the state vector. Physical states are represented by state vectors with a radius less than one. If each of the expectation values can only be estimated to within some finite accuracy, it is possible that Eq.(65) will yield an unphysical density matrix. What, if anything, can be concluded from this unphysical density matrix? The metric of accuracy of tomography, fidelity, does not yield meaningful results for unphysical density matrices (for example, fidelities greater than one can be obtained). In the next section, I will show how to confine the estimate produced by tomography to a physical state in a meaningful way.

In general, we represent our quantum state as a density matrix which is a Hermitian matrix $X$ of size $2^q \times 2^q$. The quantum tomography measurement process collects measurements

$$\mu = A(\rho) + n, \qquad (A(\rho))_i = Tr(\hat{O}_i^* \rho) \qquad (69)$$

where $n$ is a term that represents noise, and $\hat{O}_i^*$ is a Hermitian matrix that represents an observable. For experimental reasons, $\hat{O}_i^*$ is the tensor product of Pauli matrices [139]. Note the similarity between Equations 50, 52 and 69, that is to say, the model of noise associated to quantum measurement.

## 5 Experiments and simulations

Of all the current methods of Quantum State Tomography (QuST) [100-102,132,139-143], that is to say, Maximum Likehood, Estimation via Backprojections, Linear Regression Estimation, Compressed Sensing, and Kalman Filter Approach, we will use in here the last one, because, it is more robust in the presence of noise [143].

On the other hand, in this Section, it is vital importance the reference that gives us the QuST which it is estimated later than the quantum process which uses some internal representation of the image such as FRQI, NEQR, and QuBoIP (or simply: QuBo). The use of the QuST will give us a realistic notion of how the quantum measurement noise affects the quality of the outcome according to the internal representation of the selected image.

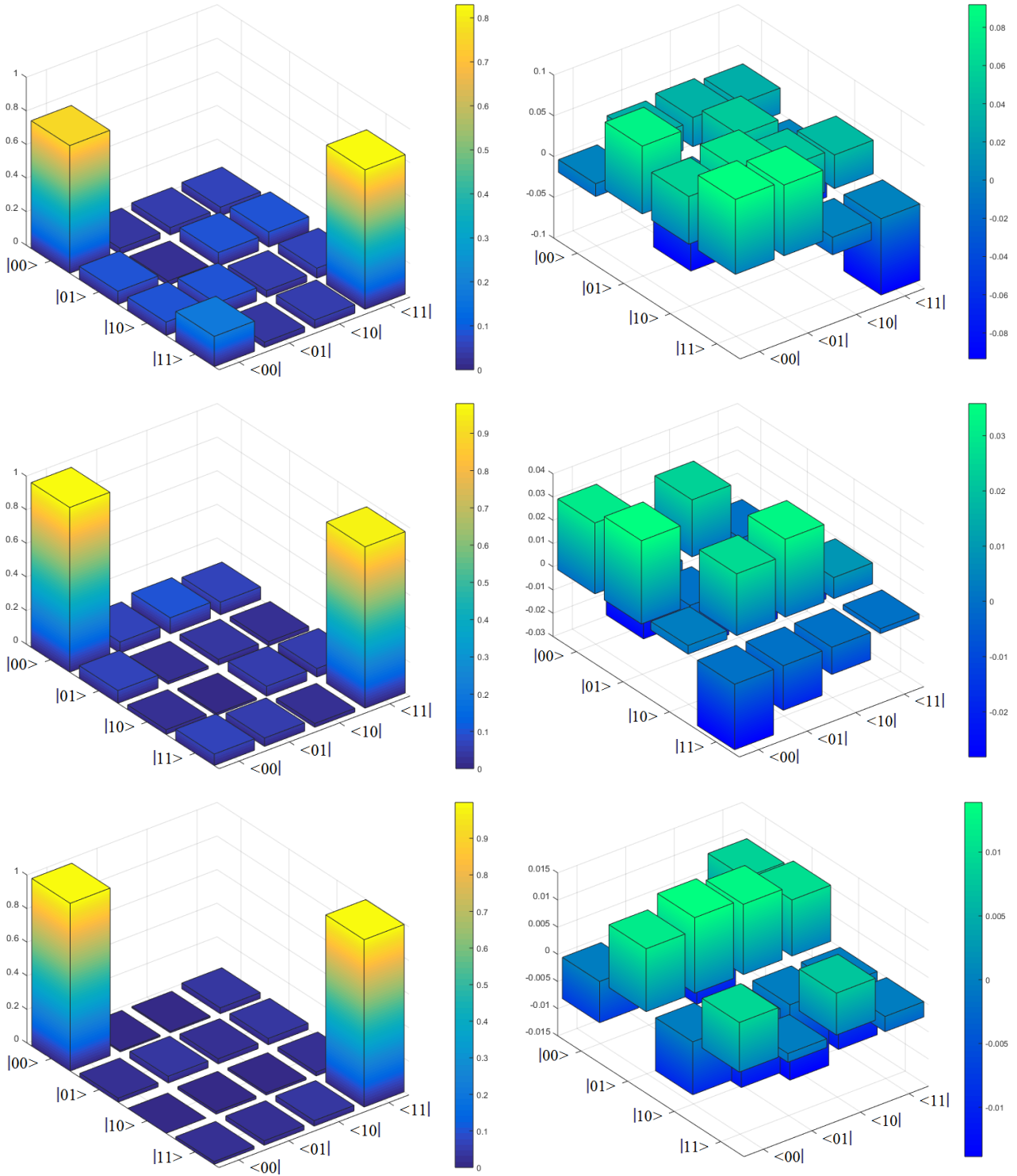

**Fig. 18** Quantum state tomography for 2 qubits, after a quantum image process which involves a particular internal image representation, namely, top-left: real-part for FRQI, top-right: imaginary-part for FRQI, middle-left: real-part for NEQR, middle-right: imaginary-part for NEQR, bottom-left: real-part for QuBo, and finally, bottom-right: imaginary-part for QuBo. In real part, and coordinates $|00\rangle, \langle 00|$, and $|11\rangle, \langle 11|$, both should give an exact 1. On the other hand, NEQR values were normalized to 1, to equate the 3 techniques.

Figure 18 shows us real and imaginary part of QuST after the use of FRQI, NEQR and QuBo. The other techniques for internal representation of the image (named in Section 4) are not representative, because they are in the range between FRQI and NEQR, and all are worse outcome than QuBo.

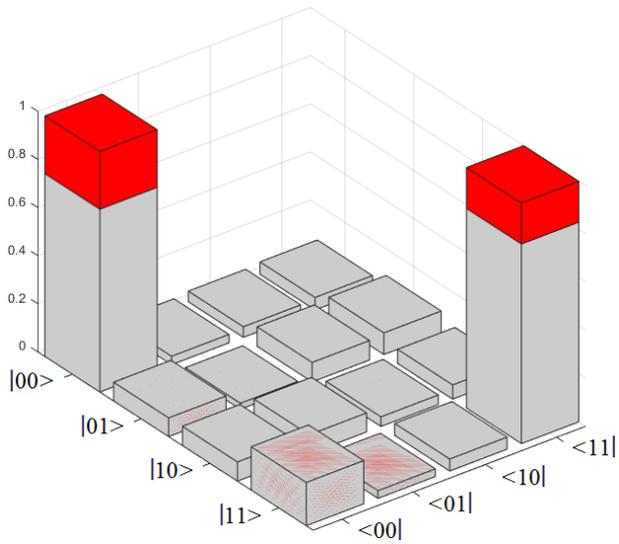 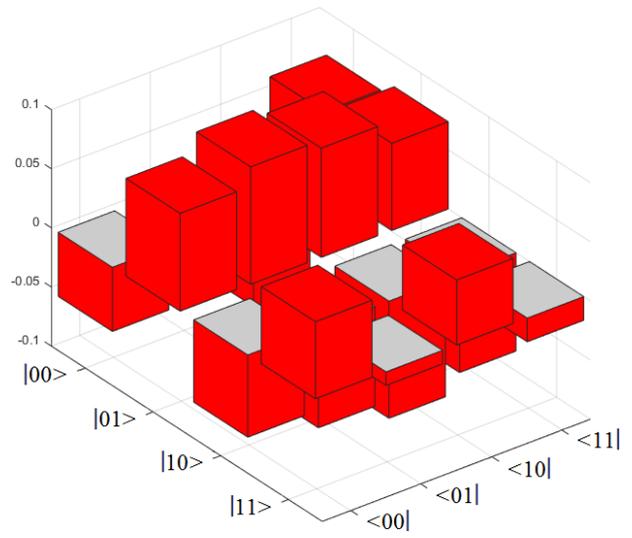
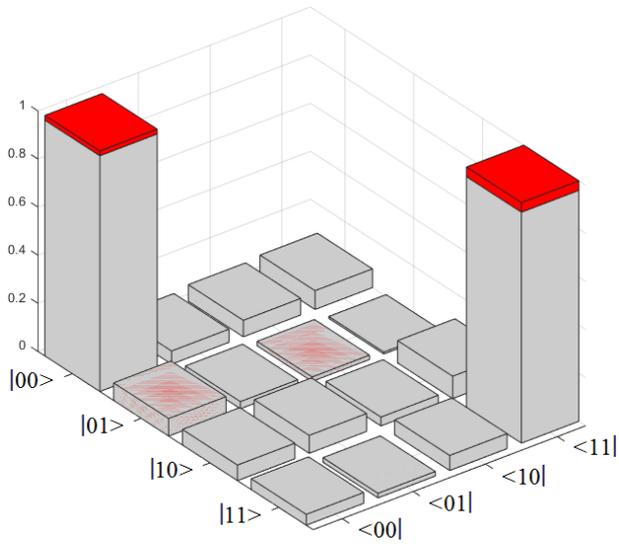 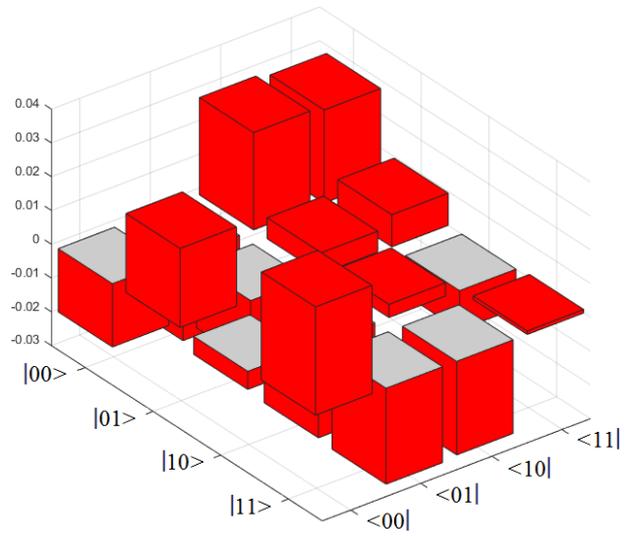
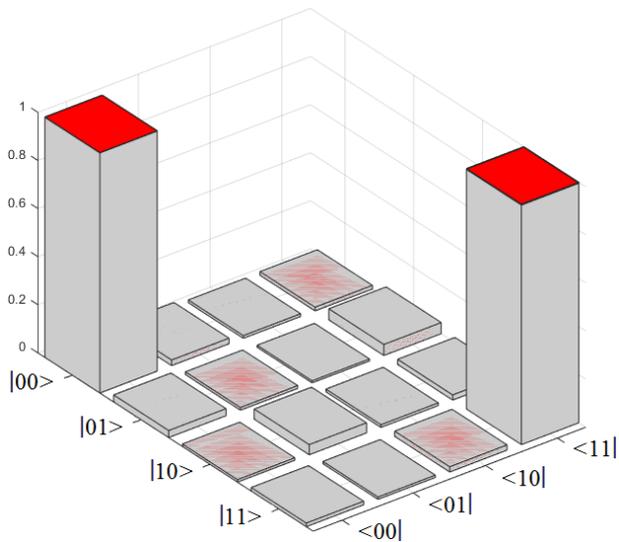 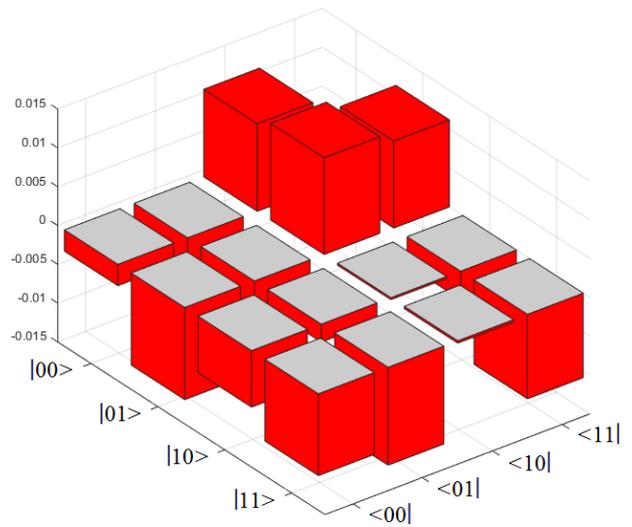

**Fig. 19** With the same approach and the same order of the previous figure, the graph shows the measured values in gray and the measurement errors in red, that is to say, from left to right: real and imaginary part, and, from top to bottom: FRQI, NEQR, and QuBo.

For the purpose of having a quantitative and comparative view of measurement error states according to the technique of internal representation of the image, we can see Figures 19 and 20, which bring out such errors in relation to the value of he considered states. Both graphs show clearly that the biggest errors are introduced by FRQI. In particular and for the three cases, the errors increases as the number of considered qubits.

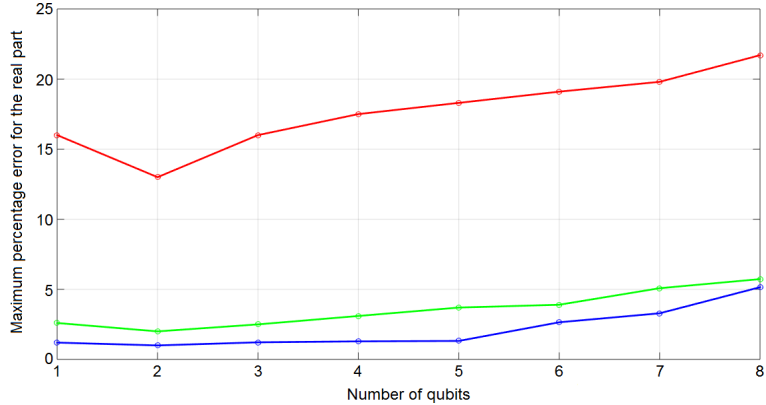

**Fig. 20** Maximum percentage error (for the real-part) in term of number of qubits, with FRQI in red, NEQR in green, and QuBo in blue.

TABLE II
MAXIMUM PERCENTAGE ERROR IN TERM OF INTERNAL IMAGE REPRESENTATION FOR 2 QUBITS.

| internal image representation | maximum percentage error for real part (estimated) |
|---|---|
| FRQI | 13 |
| NEQR | 2 |
| QuBo | 1 |

TABLE III
TIME COMPLEXITY AND IMAGE TYPE IN TERM OF INTERNAL IMAGE REPRESENTATION FOR $2^n \times 2^n$ IMAGES.

| internal image representation | time complexity | image type |
|---|---|---|
| FRQI | $O(2^{4n})$ | colour |
| NEQR | $O(qn2^{2n})$ | gray |
| QuBo | $O(2^{2n})$ and $O(32^{2n})$ | gray and colour (respectively) |

Tables II and III show us: maximum percentage error, and time complexity with a description of the image type, respectively. As we can see, QuBo has the slightest error, with the slightest time complexity, and although errors of NEQR and QuBo are in the same order, the first one serves to gray images, exclusively, while the latter can be applied in an unrestricted form to gray and colors images.

## 6 Conclusions

In this paper, we try to establish the impact of quantum measurement on the outcomes caused by the most used techniques of internal representation of an image involved in a quantum process. The results of the simulations of Section 5 indicate that FRQI gives the worst results, while NEQR and QuBo give the best results. However, while NEQR is only for images in gray, QuBo is equally useful for both types of images, that is to say, gray and color. This is because FRQI is alarmingly sensitive to quantum measurement. Then, FRQI was the worst method of internal representation of the image for its outrageous sensitivity to quantum measurement. The experiments were repeated numerous times and the results were always the same.

In another vein, Quantum Image Processing is a young discipline with a great potential, which (among other things) must deal with a number of problems, for example, if we call *the method* to the procedure consisting of:
   a) classical-to-quantum interface (Cl2Qu) composed mainly by the qubits preparation
   b) quantum algorithm (those used in the literature for QuIP)
   c) quantum-to-classical interface (Qu2Cl) composed mainly by the quantum measurement,

such a method should respect elementary requirements of DIP, such as
1) <u>Geometric Invariance</u>: The method should preserve the geometry and relative sizes of objects in an image. That is, the subject matter should not change under their application.
2) <u>Contrast Invariance</u>: The method should preserve the luminance values of objects in an image and the overall contrast of the image.
3) <u>Noise</u>: The method should not add noise or other artifacts to the image, such as ringing artifacts near the boundaries.
4) <u>Edge Preservation</u>: The method should preserve edges and boundaries, sharpening them where possible.
5) <u>Aliasing</u>: The method should not produce jagged or "staircase" edges. It is very likely that this will happen for internal representations of the image of the type FRQI and NEQR (among others).
6) <u>Texture Preservation</u>: The method should not blur or smooth textured regions.
7) Over-smoothing: The method should not produce undesirable piecewise constant or blocky regions.
8) <u>Application Awareness</u>: The method should produce results appropriate to the type of image and order of resolution. For example, the interpolated results should appear realistic for photographic images, but for medical images the results should have crisp edges and high contrast. If the interpolation is for general images, the method should be independent of the type of image.
9) <u>Sensitivity to Parameters</u>: The method should not be too sensitive to internal parameters that may vary from image to image.

Any article of QuIP that works with generic qubits analyzes these themes. Still less, almost none of them compares its results with its classical counterpart, only a few, e.g., QuBoIP [50].

In contrast, the most of papers on QuIP, crucially, the only thing potentially new such papers have done is write down a few equations. However, there is no explanation for what these symbols mean nor what they achieve. Then the papers ends. Usually, the authors has simply taken the classical image processing equations and replaced some of the variables with quantum states and others with quantum operations and assumes it solves an analogous problem.

Generally, it reads as series of disconnected "facts" about a range of topics in quantum information. Many of these facts are misstated or misinterpreted. This is not meant to be snide, but it reads similar to a poorly written term paper.

Besides, these topics bring some questions:
- What kind of classical-to-quantum and quantum-to-classical interfaces do authors propose?
- How do the authors prepare the qubits? As part of the classical-to-quantum interface.
- What kind of quantum measurement do authors use? Weak alone, or weak and strong? May be, state tomography? As part of the quantum-to-classical interface.
- What's the impact of quantum measurements on the outcome (recovered images)?
- Why the authors use FRQI and NEQR as a representation of the image? We know that FRQI is extremely sensitive to quantum measurement, and NEQR only works with gray images.
- How does the proposed procedure in presence of noise? Is it robust?
- Why the authors did not use comparative metrics (Mean Absolute Error, Mean Square Error, Peak Signal-to-Noise Ratio, etc) between the original and recovered images? In fact, it lacks a metric to measure and compare with traditional techniques, i.e., classical techniques.
- Among others.

However, in any work are answered those questions.

In fact, the literature shows a lack of knowledge about a minimal notion of the concepts of algorithmic robustness (in the presence of noise), stochastic processes, and Theory of Measurement, since or not treated, or they are treated as secondary things.

On the other hand, there are things that are permissible in certain quantum algorithms (especially if they work with CBS), but not in QuSP or QuIP with generic qubits.

It is important to mention at this point that to a quantum computer today would not running several days to resolve a brute-force calculation, but that it must solve the problem almost instantly because the decoherence carried out a sabotage in quantum computers that are not adiabatic. If to this is added the problem of quantum measurement when we work with generic qubits, the recommendation is clear, if you can, please, work with CBS almost exclusively. QuIP is a passive victim of this problem. Besides, there is no practice quantum computer today and still less an adiabatic. Therefore, how is it possible that the QuIP ignores these disadvantages if the quantum computer does not do so?

Of course, these are qualitative, quantitative and irrefutable objections. However, some work in the literature analyzed this? Obviously not. Then, what is the conclusion? Is QuIP a discipline with better attributes than Quantum Information Processing? Absurd.

The situation would be very different if we could use a reusable quantum Turing machine to which we could transfer aspects of implementation of an algorithm directly to the hardware, but this is not the case of the current QuIP.

Another issue of primary importance to the implementation of the quantum algorithms of QuIP corresponds to the fact that Figures 4 through 7 are for each pixel and not for the full picture, given that, in which quantum computer today comes a complete image at the same time? In that work this is clarified?

In another order of things and as it has been clarified in due course, the use of classical equations with the notation of Dirac, does not convert directly to a classical algorithm (coded in MATLAB® [49]) in one quantum algorithm for a realistic and practical deployment of QuIP. As a direct result, these commentaries produces a very simple question: why doesn't developed Quantum Signal Processing [13, 14], if it has existed for 15 years? It is clear that the reason is that something affects the recovered signal, that is quantum measurement, which introduces a noise, and we don't know the statistics of the noise. Remember that, the same act of measure collapses the wave function, because the quantum measurement is a disturbance which is manifested as a noise. This is accompanied by: noise of state (which comes with the classic original image, or is due to a faulty preparation of the qubits), decoherence, etc.

What we have said so here is the inevitable penalty that receives the one who is irreverent with the Quantum Measurement.

Newly, Quantum Image Processing is a young discipline with a great potential, and at this point, the choice of title for this paper was not intended to denigrate this novel discipline, but on the contrary, both the author and the rest of the scientific community we are waiting for a series of papers to equate the performance of QuIP with DIP [45-48] in filtering, enhancement, restoration, compression and super-resolution of images, for the purpose of QuIP becomes a real alternative with its own identity.

Finally, the classical simulations were implemented in MATLAB® R2015a (Mathworks, Natick, MA) [49] on a notebook with Intel® Core(TM) i7-4702 MQ CPU 2.20 GHz and 8 GB RAM on Microsoft® Windows 7© Ultimate 64 bits. While, the quantum experiments were implemented in laboratories of National Commission of Atomic Energy with quantum optics facilities.

**Acknowledgments** M. Mastriani wishes to thank all the technical staff of the various laboratories of the National Commission of Atomic Energy for the help they gave me in the preparation of experiments. It is impossible to name them all here, simply, thank you all for all.